\documentclass[%
 reprint,
 amsmath,amssymb,
 aps,
]{revtex4-2}

\usepackage{float}
\usepackage{graphicx}
\usepackage{dcolumn}
\usepackage{bm}
\usepackage{placeins} 
\usepackage{color}
\usepackage{subfigure}

\begin{document}

\title{Models of ultra-heavy dark matter visible to macroscopic mechanical sensing arrays}%

\author{Carlos Blanco$^{a,b}$}
\email{carlos.blanco@fysik.su.se, ORCID: orcid.org/0000-0001-8971-834X}

\author{Bahaa Elshimy$^{c}$}

\author{Rafael F. Lang$^{c}$}

\author{Robert Orlando$^{c}$}

\affiliation{${}^a$Department of Physics, Princeton University, Princeton, NJ USA}
\affiliation{${}^b$The Oskar Klein Centre, Department of Physics, Stockholm University, AlbaNova, SE-10691 Stockholm, Sweden}
\affiliation{${}^c$Department of Physics, Purdue University, West Lafayette, IN USA}

\date{\today}%

\begin{abstract}
In recent years, the sensitivity of opto-mechanical force sensors has improved leading to increased interest in using these devices as particle detectors. In this study we consider scenarios where dark matter with mass close to the Planck scale may be probed by a large array of opto-mechanical accelerometers. We motivate a macroscopic mechanical search for ultra-heavy dark matter, exemplified by the efforts of the Windchime collaboration, by providing a survey of the model space that would be visible to such a search. We consider two classes of models, one that invokes a new long-range unscreened force and another that is only gravitationally interacting. We identify significant regions of well-motivated, potentially visible parameter space for versatile models such as Q-balls, composite dark matter, relics of gravitational singularities, and gravitationally produced ultra-heavy particles.
\end{abstract}

\maketitle


\section{Introduction and Motivation}
Opto-mechanical detectors are sensitive to a previously unprobed region of dark matter (DM) parameter space \cite{Carney:2020xol}. In general, an array of macroscopic accelerometers can be sensitive to heavy DM interacting through a long range force \cite{Carney:2019pza}. Specifically, a DM candidate which acts like a mesoscopic (ng to $\mu$g-scale) compact particle coupling to some unscreened charge of ordinary matter can leave detectable track-like signatures in these detectors. Recently, there has been an increase of interest in using force sensors to search for Beyond the Standard Model (BSM) physics ~\cite{Carney:2019pza,Carney:2020xol,Moore:2020awi,Monteiro:2020wcb, Ghosh:2019rsc,Kawasaki:2018xak,Hall:2016usm,Manley:2019vxy,Manley:2020mjq}. The pioneering Windchime experiment, modeled after the proposal detailed in \cite{Carney:2019pza}, is a prime example of this effort. The advent of new direct detection methods motivates a closer inspection of the models occupying the mesoscale parameter space and urges an investigation into the sensitivity of impulse-sensing experiments to DM candidates predicted by these models.

The goal of this study is to explore the range of models that a large three-dimensional array of force sensors might be sensitive to in the context of Planck-scale DM. As the mesoscale particle travels through the detector volume, it imparts a series of small correlated impulses on the sensors closest to its trajectory. By monitoring the dynamics of macroscopic oscillators, one gains directional information and can robustly reject backgrounds that are relevant to other techniques ~\cite{Carney:2019pza}. The sensitivity of such an approach is determined by the physical parameters of the array as well as by the mass and noise inherent in the individual sensors. Here, we consider the fundamental noise floor set by thermal and quantum noise. We discuss an idealistic detector design that, while beyond immediate experimental capabilities, represents a possible ultimate goal and direction for the Windchime collaboration.

There have previously been searches and proposals for the detection of electrically-charged DM candidates at the Planck scale ($10^{16}\;\text{GeV} \lesssim M_{\chi} \lesssim 10^{22}\;\text{GeV}$) \cite{lehmann_johnson_profumo_schwemberger_2019, Meissner:2018cay}. Here, we examine electrically neutral Planck mass DM that interacts either through a new long-range force or simply through gravity. We consider the search for track-like impulse signatures of these DM candidates. When the dark matter-nucleon cross section is much less than the geometric cross section of the dark matter state, existing searches for Planck-scale DM become blind to the collision observables. Previous searches for such heavy DM have taken place at large detectors directly \cite{Arafune:2000yv, Piotrowski:2020ftp, Mayotte:2016ufk, Pshirkov:2015cca, SLIM:2008bwg} or through observation of energy loss signatures within neutrino detectors \cite{Ahlers:2018mkf} and resonant-bar gravitational wave detectors \cite{ROG:2015teb} as well as by looking for characteristic defects left within ancient target materials such as mica or meteorites \cite{Liu:1988wn, Price:1988ge}. Fifth-force searches \cite{Schlamminger:2007ht,Wagner:2012ui,Abbott:2021npy} and existing bounds on DM self-interaction \cite{Robertson:2016xjh} also provide stringent constraints on this parameter space.

In this study we intend to provide a non-exhaustive survey of the possible Planck-scale DM candidates which can be probed using large arrays of accelerometers. While we mention suggested production mechanisms for the DM candidates in the model classes we examine, we do not consider these in the context of constraints and instead focus on the direct detection features of the model.

We first discuss the general sensitivity of accelerometer arrays and motivate the types of models which could be explored. We consider two broad classes of models, those in which the DM interacts with the Standard Model (SM) through a new unscreened long-range force, and those in which the DM interacts with the SM only through gravity. We find that models with a new force require a significant amount of the new charge to be confined into the point-like DM and discuss two generic and complementary model classes that can achieve this. In the first case, such charge confinement is realized via the formation of bound states of constituent fundamental particles. These constituents may be SM particles \cite{Bodmer:1971we,Witten:1984rs} or be part of a dark sector \cite{Bai:2018dxf,Hardy:2014mqa,Hardy:2015boa,Gresham:2017zqi,Redi:2018muu,Detmold:2014qqa,Detmold:2014kba,Gresham:2018anj,Krnjaic:2014xza,Gresham:2017cvl}. In the second case, charge is confined into states that are non-fundamental particles; in specific, non-topological solitons. We discuss these solutions to the field equations in the context of three types of Q-balls \cite{Enqvist:2001jd, cotner_kusenko_2016}, and present a simplified unifying family of models. 

In the case where the DM interacts with the SM only gravitationally, the parameter space can be split into two regions, sub-Planckian and super-Planckian DM. The DM candidate is only expected to be a fundamental particle when its mass is below the Planck mass. Since the only parameter visible in these models is the DM mass, the model classes are limited only by production mechanisms, which we do not aim to constrain here. Therefore, we simply discuss models which can generate DM at these scales. We discuss sub-Planckian candidates in the context of WIMPZILLAs \cite{Chung:1998zb, Chung:1998ua}, though we note that it is increasingly difficult to produce such particles within two orders of magnitude of the Planck mass since the reheating temperature can only be pushed up to inflationary scales. We identify the Planck-scale relics of extremal black holes \cite{Bernal:2020bjf, lehmann_johnson_profumo_schwemberger_2019, Bai:2019zcd, Aydemir:2020xfd, Abbott:2021npy,MacGibbon:1987my,Salvio:2019llz} as ideal DM candidates to look for with this approach. These relics may be produced above the Planck mass and therefore bracket the second region of parameter space. We finally note that, indeed, mechanical arrays are ideally suited to look for these relics and, as of writing this document, are the only proposed type of detector to do so.

The paper is structured as follows. In section \ref{sec:sec2}, we describe the strategy of heavy DM searches with an array of accelerometers and describe the general features of visible models. In section \ref{sec:sec3} we discuss the model space for the new-force case and gravitational coupling case. Finally, in section \ref{sec:sec4}, we discuss the regions of parameter space which could be explored with this strategy and conclude.

\section{Detection of heavy DM with opto-mechanical oscillators}
\label{sec:sec2}
Arrays of macroscopic accelerometers can be used to probe long-range interactions between heavy dark matter (DM) and the standard model (SM).  Generically, large DM masses and long-range interactions are needed in order to observe a given model more readily. Planck-scale dark matter with gravitational couplings is a benchmark model which could be seen with large arrays instrumented with billions of sensors and with spacings of cm to mm~\cite{Carney:2019pza}. As we will discuss, for smaller arrays (i.e. fewer sensors), larger couplings or heavier DM masses are needed in order to acquire a significant signal-to-noise ratio. Therefore, we focus on models which predict dark matter candidates with mass $m_{DM} > M_{P}\times 10^{-3}$, where the Planck mass is $M_{P}= 1.2 \times 10^{19}\; \text{GeV} = 21.8 \; \mu \text{g}$. 

Here, we will consider two types of forces, gravity and a long-range dark force. In the case of a dark force, we invoke a light boson as the force mediator, generating a potential given by
\begin{equation}
    V \sim \frac{g_\phi^2 Q_{SM} Q_{DM}}{ r} e^{-m_{\phi}r},
    \label{eq:yukawa}
\end{equation}
where $m_{\phi}$ is the mass of the light mediator which mediates the DM-SM interaction, $g_{\phi}$ is the coupling constant for the interaction mechanism, $Q_{DM}$ and $Q_{SM}$ are the charges of the interacting DM and SM particles under this interaction, and $r$ is the interaction distance between the DM and the SM charge. This generic Yukawa potential has a characteristic length $r_{\phi} = m_\phi^{-1} \approx 0.2 \;\mu \text{m} (\text{eV}/m_\phi)$. Thus, at ranges within about a centimeter and mediator masses lighter than $\approx 1\; \mu \text{eV}$, the interaction is essentially Coulombic, i.e. giving a force,
\begin{equation}
    F \sim \frac{g_\phi^2 Q_{SM} Q_{DM}}{r^2} \sim \frac{G m_{\mathrm{sensor}} m_{DM}}{r^2},
\end{equation}
where $G$ is given by 
\begin{equation}
    G=\frac{g_{\phi}^2 \lambda_{SM} \lambda_{DM}}{4 \pi}.
\end{equation}
Here, $g_{\phi}$ is the dark-force coupling and $\lambda_{SM(DM)}$ is the SM (DM) charge-to-mass ratio. The gravitational case is recovered when $g^2_{\phi}\lambda_{SM}\lambda_{DM}=(4\pi M_{P}^2)^{-1}$. Despite the Coulombic analogy, the charges discussed in this study belong to a new dark-gauged sector that are independent and different from the SM electric charge.

Following the discussion in ref.~\cite{Carney:2019pza}, when a dark matter particle of charge $m_{DM}$ passes by a detecting mass of charge $m_{\mathrm{sensor}}$ at a distance of closest approach $b$, the transverse component of the force is given by
\begin{equation}
    F_{\perp}=\frac{G m_{\mathrm{sensor}} m_{DM} b}{(b^{2}+v^2 t^2)^{3/2}} , \label{eqn:Fperp}
\end{equation}
where  $v \approx 220\; \text{km/s}$ is the mean DM velocity~\cite{Evans:2018bqy}.
Note that for a densely packed array, the point of closest approach is essentially the same as the impact parameter. From here on, we consider $b$ as the impact parameter. 

The force felt by a sensor is given by
\begin{equation}
    F(t)= F_{int}(t)+F_{th}(t)+F_{meas}(t),
\end{equation}
where $F_{th}$ is the thermal noise, $F_{meas}$ is the noise induced via measurement, and $F_{int} = F_\perp$ is the force due to the interaction with DM as in Eq.~\ref{eqn:Fperp}. $F_{meas}$ is bounded from below by the quantum measurement limit and $F_{th}$ is characteristic of Brownian motion and is dependent on the physical characteristics of the individual sensors. The physical observable in these types of detection schemes is the impulse imparted on the accelerometer by a passing DM particle, given by
\begin{equation}
    I = \overline{F}_\perp\tau,
\end{equation}
where $\tau \sim b/v \approx 10\; \text{ns}\;(b/1\;\text{mm})$ is the duration of the interaction, and $\overline{F}_\perp$ is the average force over that time given by
\begin{align}
    \overline{F}_\perp = \frac{2Gm_{DM} m_{\mathrm{sensor}}}{b^2}.
\end{align}

Given a thermal noise which is proportional to the square root of measurement time,
\begin{equation}
    \Delta I^2 = \alpha t_{meas},
\end{equation}
the signal-to-noise ratio ($\mathrm{SNR}$) of an interaction across $N \sim L/b$ such sensors is given by
\begin{equation}
    \mathrm{SNR}^2 = \frac{N\overline{F}_\perp^2 \tau}{\alpha},
\end{equation}
when the measurement is done for a time during the interaction $t_{meas} \approx \tau$ and where $\alpha$ is the proportionality constant relating the measurement time to the noise inherent in an impulse measurement.

Assuming that the measurement-induced noise is subdominant to the thermal noise of the accelerometer, a mechanically coupled sensor at temperature $T$ has a characteristic $\alpha$ given by the following,
\begin{equation}
    \alpha_{mech} \approx 4 m_{\mathrm{sensor}} k_B T \gamma,
\end{equation}
where $\gamma$ is the sensor's mechanical  damping rate. 

In an array of side length $L$ where the sensors are separated by a spacing $d\sim 2 b$, the $\mathrm{SNR}$ is then given by
\begin{align}
    \mathrm{SNR}^2&=\left( \frac{g_\phi^2 \lambda_{SM} \lambda_{DM} hc}{4\pi} \right)^2 \frac{m_{\mathrm{sensor}} m_{DM}^2 L}{d^4 v k_b T \gamma} \nonumber \\
    &=\frac{G_N^2 m_{\mathrm{sensor}} m_{DM}^2 L}{d^4 v k_b T \gamma} \nonumber \\
    &\approx 10^2 \left(\frac{m_{DM}}{1\; \text{mg}}\right)^2 \left(\frac{m_{\mathrm{sensor}}}{1\; \text{mg}}\right) \left(\frac{1 \; \text{mm}}{d}\right)^4,
    \label{eq:snr1}
\end{align}
where we take an experiment like the one discussed in Ref.~\cite{Carney:2019pza} with a side length $L=1\;\text{m}$, cryogenic dilution-fridge temperature $T=10\;\text{mK}$, and mechanical damping rate $\gamma =10^{-8}\;\text{Hz}$~\cite{Carney:2020xol}. In the second line of Eq.\ref{eq:snr1} we have used the gravitational coupling limit. In order to make a statistically significant detection, the signal-to-noise ratio must be greater than 5 after correcting for the false-positive rate. An SNR of 5 corresponds approximately to a 5$\sigma$ event assuming uncorrelated noise between the sensor masses (i.e. $\mathrm{SNR} \approx \mu/\sigma$). This serves as a minimum benchmark for the statistical significance of an event which we will call visible. 

\begin{figure}
\includegraphics[width=1\linewidth]{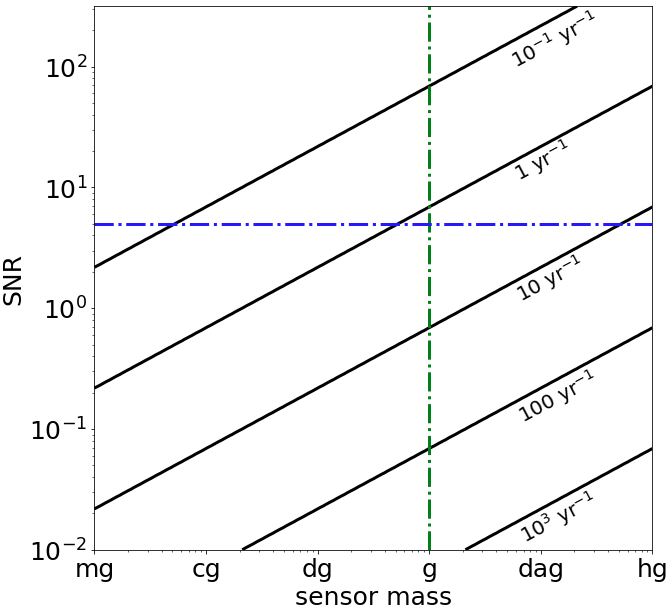}
\includegraphics[width=1\linewidth]{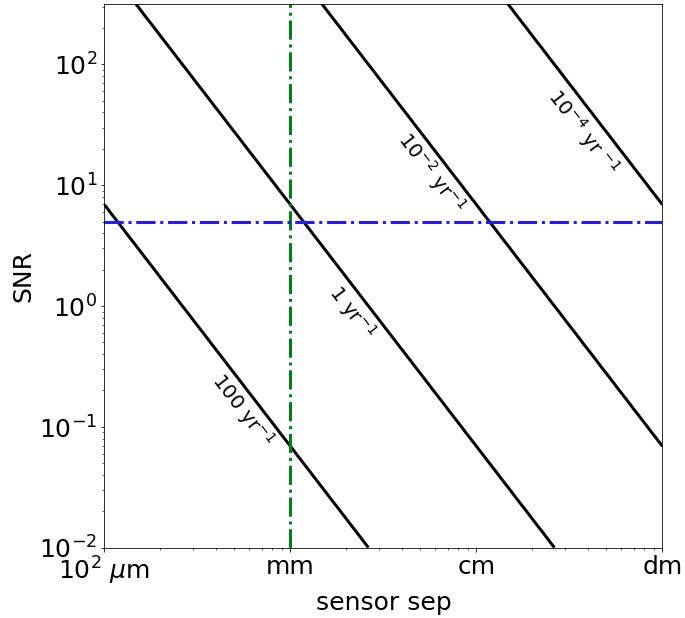}
\caption{The top frame shows the dependence of the $\mathrm{SNR}$ with respect to the sensor mass along contours of constant event rates while the bottom frame shows the dependence of the $\mathrm{SNR}$ with respect to the sensor separation along contours of constant event rates. In the top frame, sensor separation is 1\;mm and in the bottom frame the sensor mass is 1\;g, both for a dark matter mass of $M_p$. The SNR of 5 is also highlighted in both frames.}
\label{fig:sensorsepmass}
\end{figure}

Searches for long-range 5th forces have set stringent constraints on the coupling $g_\phi$~\cite{Schlamminger:2007ht,Wagner:2012ui,Abbott:2021npy}. From these searches, we see that, for a force mediator with $r_{\phi} \sim 1 \; \text{cm}$, the constraint on a coupling to baryon number (i.e. $\lambda_{SM} = N_A = 1/\mathrm{proton}\; =\;1\; \text{GeV}^{-1} = 6.02\times10^{23}\; \text{g}^{-1}$) is $g_\phi^2 \lesssim 2\times10^{-42}$. From these constraints we can conclude that if the coupling to the SM is proportional to the baryon content of ordinary matter, which is common for un-screened forces, then the dark matter charge-to-mass ratio must be at least
\begin{align}
\lambda_{DM} &\gtrsim 6.5\times10^{33} \; \text{g}^{-1} \nonumber \\ 
&\sim 1200\; \text{GeV}^{-1}    \left(\frac{T}{10\text{mK}} \right)^{1/2}  \left(\frac{\gamma}{10^{-8}\text{Hz}} \right)^{1/2} \label{eq:minLambda}
\end{align}

in order for a transit through the detector array to have a $\mathrm{SNR}$ greater than 5. Fig.~\ref{fig:sensorsepmass} shows the dependence of the signal-to-noise ratio with respect to sensor mass and separation, along with contours of expected event rates. As expected, instrumenting the array with sensors towards the gram scale and sensor separation at or below about a millimeter would drastically improve sensitivity. 



\subsection{Flux Limit}
The rate $R$ of such DM particles transiting the detector is determined by the flux through the detector:
\begin{align}
\label{eq:rate}
    R &= \frac{a \rho_{DM}vL^2}{m_{DM}} \\ \nonumber
    &\approx 1 \; \text{yr}^{-1} \left(\frac{M_{P}}{m_{DM}} \right) \left(\frac{L}{1\; \text{m}} \right)^2 \left(\frac{a}{1} \right),
\end{align}
where $\rho_{DM}=0.3\;\text{GeV}/\text{cm}^3$ is the local DM energy density and $a$ is a flux factor related to the mass fraction of the DM made up by the candidate in question defined as $a=1$ when $\rho_{candidate}=0.3\;\text{GeV}/\text{cm}^3$~\cite{Evans:2018bqy}. While $a$ is generally assumed to be unity, it is useful to parametrize situations where the DM candidate presents in either overabundance or underabundance. For example, if the DM candidate is a subcomponent of the total DM then $a<1$ while if the DM candidate makes up all the DM and there is a local overdensity then $a>1$.  We therefore have to limit our search down to dark matter candidates with masses less than the Planck mass in order to observe at least one event during a year long search with this ideal 1$\;\text{m}^3$ detector when $a=1$.


\section{Models and Relevant Parameter Space}
\label{sec:sec3}
Neutral mesoscale (ng to $\mu$g-scale) and microscale (mg) dark matter is well motivated in a variety of different models. We take a broad survey of models which predict DM with mass $10^{16}\;\text{GeV} \lesssim M_{\chi} \lesssim 10^{22}\;\text{GeV}$. We begin by discussing models where a new long-range force mediates the interaction between the DM and the SM in a phenomenologically accessible way. We consider models where the dark matter is a composite state of constituent particles carrying the new charge and which are bound by an attractive potential. These constituents may belong to the SM, as is the case for strangelets and nuclearites~\cite{Bodmer:1971we,Witten:1984rs}, or they may belong to a dark sector, as with models of dark quark nuggets or dark nuclei~\cite{Bai:2018dxf,Hardy:2014mqa,Hardy:2015boa,Gresham:2017zqi,Redi:2018muu,Detmold:2014qqa,Detmold:2014kba,Gresham:2018anj,Krnjaic:2014xza,Gresham:2017cvl}. We then examine the case where the dark charge is carried (i.e. confined) in a non-fundamental particle solution to the field equations. Such is the case of non-topological soliton (local self-reinforcing wave) solutions, so called Q-balls~\cite{Coleman:1985ki,Friedberg:1976me}. Q-balls (and non-topological solitons in general) are predicted in the context of supersymmetric extensions to the SM~\cite{Kusenko:1997si,Kusenko:1997zq,Kusenko:2006gv}  or other models with flat field-directions in their potential ~\cite{Hong:2016ict,Hong:2020est,Holdom:1987ep,Lohiya:1994pmf,Stojkovic:2001qi,Macpherson:1994wf,Holdom:1987bu,Zhitnitsky:2002qa,Ogure:2002hv} such as models with scalar fields carrying a conserved-U(1) charge. These charge-carrying field configurations can be produced through phase transitions~\cite{Frieman:1988ut} in the early universe via the fragmentation of scalar condensates and could be a by-product of baryogenesis~\cite{Affleck:1984fy,Dine:2003ax,Krylov:2013qe}. 

We then move on to scenarios where the DM is only accessible to our detector through its gravitational interaction. We need only consider the mass of such DM candidates but introduce these objects in the context of Planck/GUT-scale particles produced non-thermally in the early universe i.e. WIMPZILLAs~\cite{Kolb:1998ki,Garny:2015sjg,Kolb:2017jvz,Harigaya:2016vda,Berlin:2017ife,Chung:1998ua,Chung:1999ve,Fedderke:2014ura,Chung:1998zb,Kuzmin:1998kk, park_park_2014, chung_crotty_kolb_riotto_2001, farzinnia_kouwn_2016}, primordial black holes (PBHs)\cite{li_lu_wang_zhou_2020, bernal_zapata_2021, gondolo_sandick_haghi_2020} and the Kerr relics of PBHs \cite{lehmann_johnson_profumo_schwemberger_2019}. While these models can be made to interact with the SM through a new mediator, we note that the unscreened or fundamental charge carried by these objects is naturally small (or in the case of a global U(1) in PBHs, non-existent). Though not exhaustive, this model survey aims to span the phenomenological breadth of mesoscale DM which could be probed using arrays of opto-mechanical accelerometers.

The case of dark-charge confinement is subject to a variety of experimental probes depending on the nature of the DM candidate. Depending on the interaction properties of the mesoscale DM, particles moving non-relativistically can leave characteristic signatures of deposited energy in ancient mica and meteorites~\cite{DeRujula:1984axn,Price:1988ge} or resonant-bar gravitational wave detectors~\cite{Liu:1988wn,ROG:2015teb}. Similarly, the energy losses expected in certain models of composite mesoscale DM and Q-balls can generate tracks which may be visible in large neutrino detectors~\cite{Arafune:2000yv,Ahlers:2018mkf}, air-shower cosmic-ray detectors~\cite{Mayotte:2016ufk,Pshirkov:2015cca}, and other track detectors~\cite{Piotrowski:2020ftp,SLIM:2008bwg}. Furthermore, bounds on exotic fifth forces provide strong and quasi-model-independent constraints on models which interact through a new long-range force~\cite{Schlamminger:2007ht,Wagner:2012ui,Abbott:2021npy}. Finally, mesoscopically extended DM is subject to a generic self-interaction bound as derived from astrophysical observations~\cite{Kusenko:2001vu,Enqvist:2001jd,Spergel:1999mh}.

The arguments in the preceding section clarify what kind of massive particles we are looking for. We will consider two cases, one in which the DM interacts through a long-range dark force and another in which the DM interacts only gravitationally. In the former, we look for DM candidates in which the charge-to-mass ratio is much bigger than that of the SM. In fact, the relevant parameter space is where the dark sector confines charge in DM to densities about 10 million times greater than the SM. In the latter, it is sufficient to look for DM candidates with masses around the Planck mass.

\subsection{Charge Confinement}
In order for the DM to reach the large charge-to-mass ratios described above, the dark sector must confine huge charges into mesoscopic volumes ($ r \lesssim $ 1 mm). Here, we discuss two options; one in which constituent particles of mass $\mu_0$, form a bound state with mass $m_{DM}$ (Composite models), and one in which the total charge Q is confined into the volume through a non-topological soliton field configuration, e.g. Q-balls. These two generally span the kinds of models where large charges can be confined in small volumes. Even at larger radii, arrays of sensors are expected to be an effective way of looking for these heavy DM candidates~\cite{JacksonKimball:2017qgk}.

If the DM interacts with the SM through an un-screened long-range force which couples to baryon number or some other quantum number proportional to it, e.g. B-L charge, then the SM charge-to-mass ratio remains of order $\mathcal{O}(N_A)\sim6\times10^{23}/\mathrm{g}$. This means that the charge-to-mass ratio of the DM must be at least $\lambda_{DM}\gtrsim 1200\; \text{GeV}^{-1}$ in order for a DM transit to be visible ($\mathrm{SNR}>5$) in a detector array like the one described in Sec.~\ref{sec:sec2}.

Generically, these models  have the following physical parameters,
\begin{align}
    Q_{DM},&\; \text{the total charge of the DM} \\ \nonumber
    m_{DM},&\; \text{the total mass of the DM} \\ \nonumber
    R_{DM},&\; \text{the radius of the DM}. 
\end{align}
Since the charge in the DM is confined into a non-fundamental particle, $R_{DM}$ may be some mesoscopic scale. We restrict our discussion to DM which appears point-like to our detectors, i.e. $R_{DM}\lesssim 1 \text{mm}$.

Note that the cross section for DM-SM interactions cannot be assumed to saturate the geometric cross section. The non-relativistic DM-nucleon cross section for a Yukawa potential is given by the following,
\begin{align*}
    \sigma_{\chi n} &= \frac{16 \pi m_{n}^2 g_{\phi}^4 m_{DM}^{2} m_{SM}^{2} \lambda_{DM}^2 \lambda_{SM}^2}{m_{\phi}^2(m_{\phi}^2+4p^2)}, \nonumber \\
    &\approx 10^{-30}\;\text{cm}^2 \left(\frac{m_{DM}}{M_{P}}\right)^2\left(\frac{1\;\mu\text{eV}}{m_{\phi}}\right)^2\left(\frac{g_{\phi}}{10^{-21}}\right)^4
\end{align*}
where $m_n\sim 1 \;\text{GeV}$ is the mass of the nucleon, and $m_{\phi}$ is the mass-scale of the Yukawa interaction. Finally, $p\approx 1 \;\text{MeV}$ is the momentum of a nucleon with $v\approx 300\; \text{km/s}$ in the c.o.m. frame of the DM-SM interaction. We see that this is in fact much smaller than the geometric cross-section. 

Furthermore, over the parameter space we consider, $\sigma_{\chi n}\lesssim 7\times 10^{-21}\;\text{cm}^2$, which is smaller than the minimum detectable cross section for MACRO and ancient-mica searches \cite{Jacobs:2014yca, SinghSidhu:2019loh, SinghSidhu:2019nmh}, as well as the cross-section expected to be hazardous to biological organisms.

\subsection{Composite models}
For a composite DM model, the charge-to-mass ratio of the DM is given by the following,
\begin{align}
    \lambda_{DM}&=\frac{Q}{m_{DM}},\\ \nonumber
    &= \frac{\kappa q_{0}}{\kappa \bar{\mu}_0},\\ \nonumber
    &=\frac{q_0}{\bar{\mu}_0},
\end{align}
 where $\kappa$ is the constituent particle number in the composite DM, $q_0$ is the fundamental charge, and $\bar{\mu}_0 = \mu_{0}-E_{bind}/\kappa$ is the mass per constituent of the composite state. Here, $\mu_0$ is the mass of the constituents and $E_{bind}$ is the binding energy. We would naturally expect the fundamental dark-charge to be carried by these constituents to be the same as the fundamental dark-charge of the SM (e.g. B-L), thus $q_0 \sim \mathcal{O}(1)$. The charge-to-mass ratio is then given by,
 \begin{align}
     \lambda_{DM}\approx \frac{1}{\bar{\mu}_0}.
 \end{align}
 
 We can then rewrite Eq.~\ref{eq:snr1} in terms of the constituent particle number in composite DM, $\kappa= m_{DM}/\bar{\mu}_{0} = m_{DM} \lambda_{DM}$,
 \begin{align}
    \mathrm{SNR}^2&=\left( \frac{g_\phi^2 \lambda_{SM} hc}{4\pi} \right)^2 \frac{m_{\mathrm{sensor}} \kappa^2 L}{d^4 v k_b T \gamma} \nonumber \\
    &\approx 25\times 10^{-50} \kappa^2,
    \label{eq:snr}
\end{align}
where we have used the ideal experimental parameters of the last section. Therefore, an ideal detector would be sensitive to composite models where the constituent particle number is at least, 
\begin{align}
    \kappa \gtrsim 10^{25}.
\end{align}
For $m_{DM}$ less than the Planck mass, we have,
\begin{align}
    \bar{\mu}_{0} \lesssim 1\; \text{keV}.
\end{align}

\begin{figure}
    \includegraphics[width=3.5in]{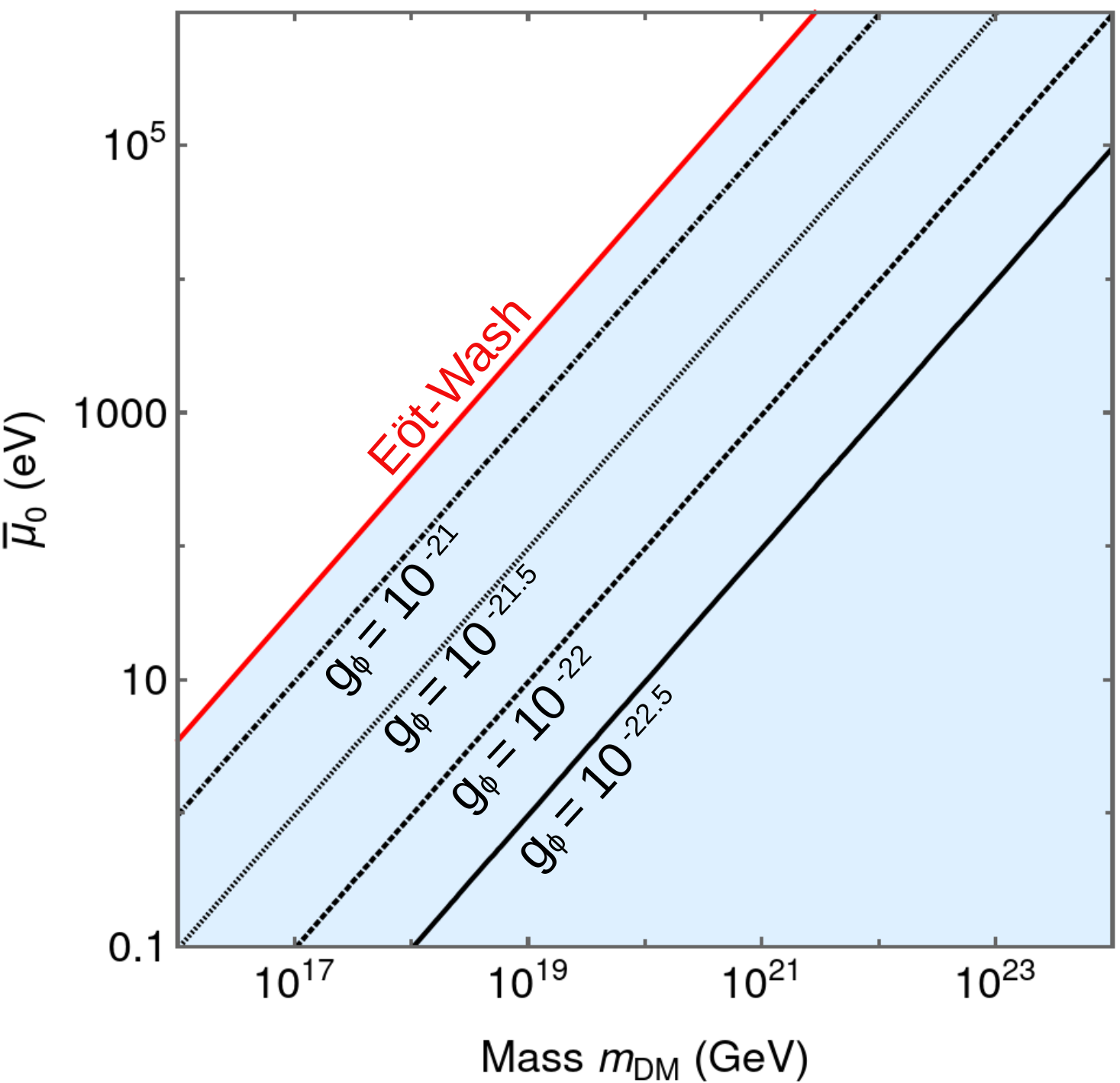}
    \caption{The blue shaded region is the projected region of sensitivity of a mechanical sensing array with characteristic length $L=1\;\text{m}$, cryogenic dilution-fridge temperature $T=10\;\text{mK}$, and mechanical damping rate $\gamma =10^{-8}\;\text{s}^{-1}$. The sensors have mass $m_{\mathrm{sensor}} = 1\;\text{mg}$ and are spaced by $d = 1\;\text{mm}$. The red line is the upper bound on $q_\phi$ from searches for long-range 5th forces $g_\phi$~\cite{Schlamminger:2007ht,Wagner:2012ui,Abbott:2021npy}.}
    \label{fig:composite}
\end{figure}

In FIG.~\ref{fig:composite} we show the region of sensitivity in the composite DM parameter space for an opto-mechanical sensing array like the one discussed in Sec.~\ref{sec:sec2}, i.e. with characteristic length $L=1\;\text{m}$, cryogenic dilution-fridge temperature $T=10\;\text{mK}$, and mechanical damping rate $\gamma =10^{-8}\;\text{s}^{-1}$. Each sensing mass is taken to be $m_{\mathrm{sensor}} = 1\;\text{mg}$. We find that when $g_{\phi}$ saturates existing bounds, that is, $g_\phi^2 = 2\times10^{-42}$, such an array is sensitive to DM masses between $10^{16}$~GeV and $10^{20}$~GeV when the constituent mass is below about $10^4$~eV.

Depending on the details of the confinement in the dark sector, models with such light constituents could be in tension with measurements of the abundances of light nuclei. In specific, these light species must be either non-thermal in origin or should be confined into non-relativistic clumps well before T $\sim$ MeV in order to not significantly affect the number of relativistic degrees of freedom during Big Bang nucleosynthesis, $\Delta N_{eff}$.

In models of composite DM, it's usual to have a dark sector in which a strongly confining interaction holds the composite state together and a light mediator which generates a long-range Yukawa potential of the form of Eq.~\ref{eq:yukawa}. 

Since the charge is held together by a second dark interaction which binds the charge tightly, we can conclude that this interaction has some gauge coupling, $g_c \gg g_\phi$. In the limit of bound state saturation, where $\kappa$ is very large and the characteristic size of the bound state scales with $\kappa^{1/3}$, the geometric radius of the composite state becomes much larger than the effective range of the strong binding force. In this limit, the cross section for velocity-independent elastic self-scattering between composite states is expected to saturate the geometric limit. The self-interaction cross section is $\sigma_{SIDM} = \xi \pi R_{DM}^2$, where $\xi$ is a model-dependent $\mathcal{O}(1)$ parameter. As discussed in e.g. Ref~\cite{Gresham:2018anj}, these interactions are subject to strong bounds, namely $\sigma_{SIDM}/m_{DM} \lesssim 1 \;\text{cm}^2 \text{g}^{-1}$. The self-interaction bound together with the $\mathrm{SNR}$ requirement can be recast as a lower bound of the rest energy per constituent, $\bar{\mu}_0$,
\begin{align}
    \sigma_{SIDM}/m_{DM} &\lesssim 1 \;\text{cm}^2 \text{g}^{-1} = 10^{-24}\; \text{cm}^2 \text{GeV}^{-1} \nonumber \\
    \frac{\xi \pi R^2}{\bar{\mu}_0 \kappa} &\lesssim 10^{-24}\; \text{cm}^2 \text{GeV}^{-1} \nonumber \\
    \bar{\mu}_0 &\gtrsim 3 \; \text{eV} \left(\frac{R}{\mu\text{m}} \right)^2 \left(\frac{10^{25}}{\kappa} \right)^2.
\end{align}
In other words, composite models with sub-micron radius and eV-scale mass per constituent could be visible to a detector like the one discussed above, while remaining well below the self-interaction bound. We would like to point out that these parameters are well-motivated by generic dark composite models. Additionally, while masses at or above the Planck scale begin to come into conflict with constraints ---depending on the thermal history of the DM--- these constraints can be relieved if the temperature at which the composite states form is significantly greater than $\bar{\mu}_0$~\cite{Gresham:2018anj}. Despite the model-dependent nature of the interactions between composite states and the SM, these interactions may be constrained with existing direct detection measurements, particularly for masses below about $10^{17}\; \text{GeV}$~\cite{Clark:2020mna,Digman:2019wdm}. We find that indeed, an opto-mechanical detector would be complementary to these existing searches insofar as it would extend the mass reach to Planckian masses and probe scenarios in which the DM-SM interaction is weak and long-range, as opposed to strong and point-like. It is worth noting that for certain long-rage mediators, the heating of interstellar clouds could also provide visible signals when the DM has a mass around or above the Planck mass~\cite{Bhoonah:2020dzs}. Such searches offer particularly interesting complementary channels.

\subsection{Q-balls}

The classification of Q-balls is generally ascribed from the behaviour of the particles' scalar field energy potentials, as described in \cite{Enqvist:2001jd, cotner_kusenko_2016}. Here, we will examine each of the three Q-ball classes introduced in these papers. The first section will explore the requirements for visible Q-balls, and then we will provide a survey of each Q-ball type in terms of its properties. We also consider a simplified Q-ball model covering the two limiting cases of Q-balls under a single description. We determine the regions of parameter space which are point-like to an array of sensors like the one discussed in Sec.~\ref{sec:sec2}. Among the limits examined will be the black hole radius and mass constraints, the established charge to mass ratio minimum, the self-interaction constraints, and the point-like requirement for a sensor array experiment. We will show that Q-balls can exist within a confined charge-to-mass region which the Windchime  experiment could provide limits for.

\subsubsection{Requirements for visible Q-balls}

The constraint that the Q-ball must be smaller than the spacing of the detector in order to appear point-like, i.e. $\sim$1mm is:
\begin{align}
    R < 1\;mm = 10^{12}\; \text{GeV}^{-1}.
\end{align}
Furthermore, the constraint that the Q-ball transit must be visible ($\mathrm{SNR}>5$), is:
\begin{align}
\frac{Q}{M} &\gtrsim 6.5\times10^{33} \; \text{g}^{-1} \\ \nonumber
&\sim 1200\; \text{GeV}^{-1}    \left(\frac{T}{10\text{mK}} \right)^{1/2}  \left(\frac{\gamma}{10^{-8}\text{Hz}} \right)^{1/2}.
\end{align}

Generically, Q-ball models feature a model-dependent scale $\varphi_0$ which is related to the characteristic vacuum expectation value of the the Q-ball field. The non-observation of new physics at the LHC implies that we shouldn't expect new non-secluded physics at scales below about 10 TeV. However, model-specific LHC observables are beyond the scope of this paper. Therefore, we cannot strictly require than $\varphi_0>10\;\mathrm{TeV}$ and instead point out regions where $\varphi_0$ may fall below 10 TeV.

Finally, we require that the radius of the Q-ball be greater than its Schwarzschild radius, $R_{sch} =2GM$. We find that in all of the relevant parameter space, this bound on the Q-ball radius is satisfied. See the appendix for a discussion of the parameter space in which the black hole condition is relevant. We note that other constraints on Q-balls exist, coming from e.g. white dwarf detonation \cite{Graham:2018efk}, at these extreme masses and charges although they do not apply in our parameter space.

\subsubsection{Simplified Q-balls}

Q-ball models can be parametrized by how their mass and radius scales with their charge, $Q$. Q-ball solutions are bracketed by the limiting cases of thick-walled and thin-walled Q-balls. Generically, the mass and radius of these non-topological solitons is given by the following,
\begin{align}
    M &= \frac{4 \pi }{3}\sqrt{2} \varphi_0 Q^p, \nonumber \\
    R &= \frac{Q^{p/3}}{\sqrt{2} \varphi_0},
\end{align}
where $\varphi_0$ is a mass scale set by the second derivative of the potential at the vacuum expectation value of the field. For a thick-walled Q-ball $p=3/4$, while for a thin-walled Q-ball $p=1$.

Immediately, we can use these parameters to restate the requirement that the charge-to-mass ratio has a minimum value, $Q/M > 1200 \text{GeV}^{-1}$, in order for the DM candidate to be visible,
\begin{align}
    \frac{Q}{M} = \frac{3R}{4 \pi} Q^{(3-4p)/3} > 1200 \text{GeV}^{-1}.
\end{align}
Similarly, since the analysis herein applies only for particles that appear point-like, the radius of the Q-balls must be smaller than the scale of the sensor separation, 
\begin{align}
    R &< d=1\text{mm}, \nonumber \\ 
    R &= \frac{4\pi Q^{4p/3}}{3M} \lesssim 5\times 10^{12} \text{GeV}^{-1}.
\end{align}
These are experiment-dependent requirements, of course.

Second, since these objects are mesoscopic, their self-interaction cross-section saturates to the geometric cross section, $ \sigma_{SIDM} = \xi \pi R^2$, where $\xi$ is an order-one number which parametrizes how effectively solid-sphere-like the Q-ball is. Here it is taken to be 1 for thin-walled Q-balls and 2 for thick-walled Q-balls~\cite{Enqvist:2001jd}. The generic self interaction bound can be expressed as follows~\cite{Kusenko:2001vu,Enqvist:2001jd,Spergel:1999mh,Robertson:2016xjh}, 
\begin{align}
    \frac{\sigma_{SIDM}}{M} &= \frac{\xi 16 \pi^3 Q^{\frac{8p}{3}}}{9M^3} \lesssim 3\times 10^4 \; \text{GeV}^{-3} , \\
    \nonumber \\
    Q &\lesssim 7\times 10^{29} \; \left( \frac{M}{M_{P}} \right)^{3/2},\;\text{Thick-walled Q-ball}, \nonumber \\
    Q &\lesssim 3\times 10^{22} \; \left( \frac{M}{M_{P}} \right)^{9/8},\;\text{Thin-walled Q-ball}.
\end{align}
Q-balls must also be stable to decay into free particles which can carry away their mass and charge. A minimal requirement for this stability is that their mass-to-charge ratio must be less than that of any charge-carrying particle, e.g.
\begin{align*}
    \frac{M}{Q} < \frac{m^{SM}_{B}}{Q^{SM}_{B}} \sim 1\; \text{GeV}, \nonumber \\
\end{align*}
for a baryon number-carrying Q-ball. Note that this is already guaranteed by the visibility condition which imposes $Q/M > 1200\;\mathrm{GeV}^{-1}$.

\begin{figure}
\includegraphics[width=1\linewidth]{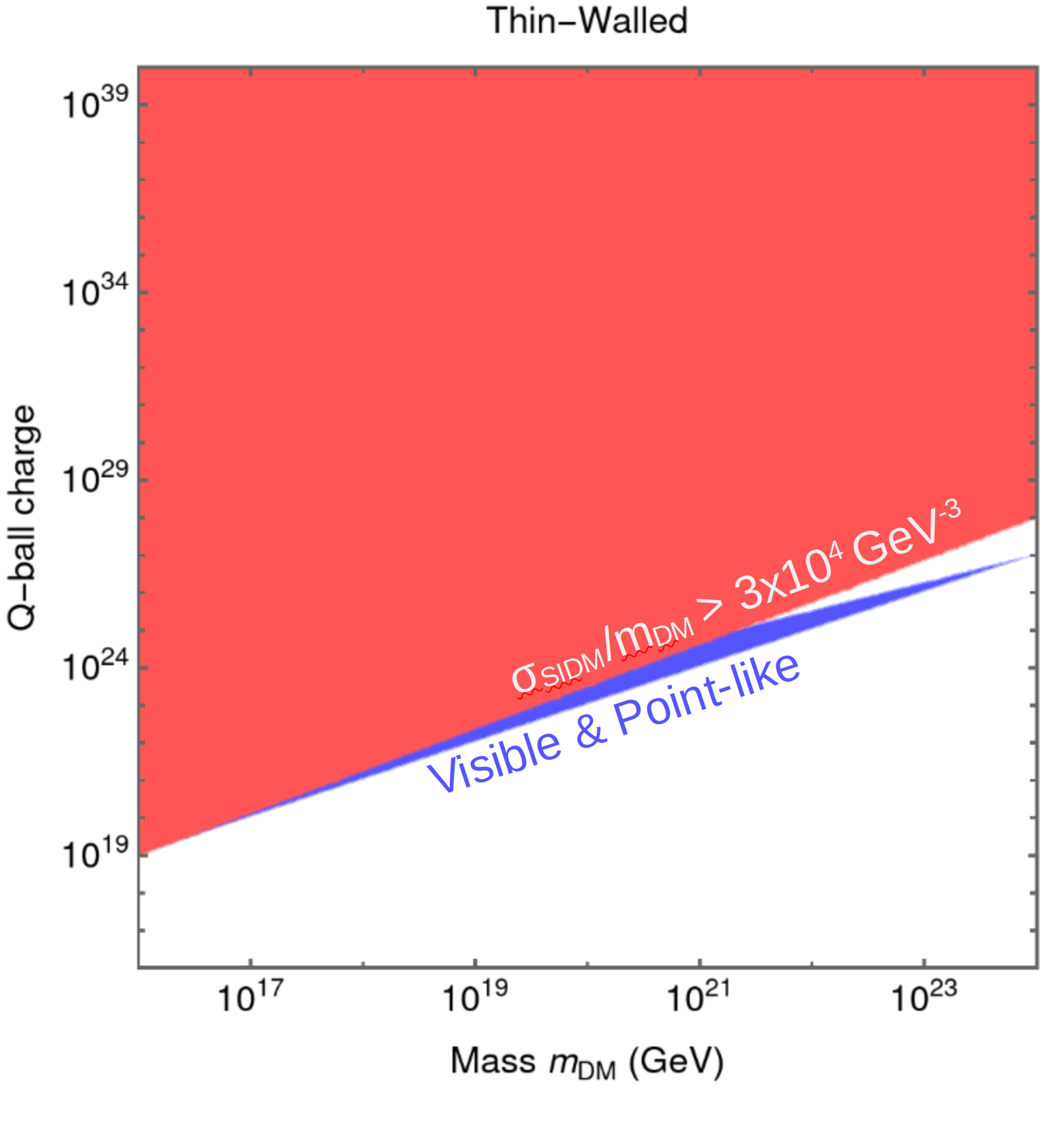}
\includegraphics[width=1\linewidth]{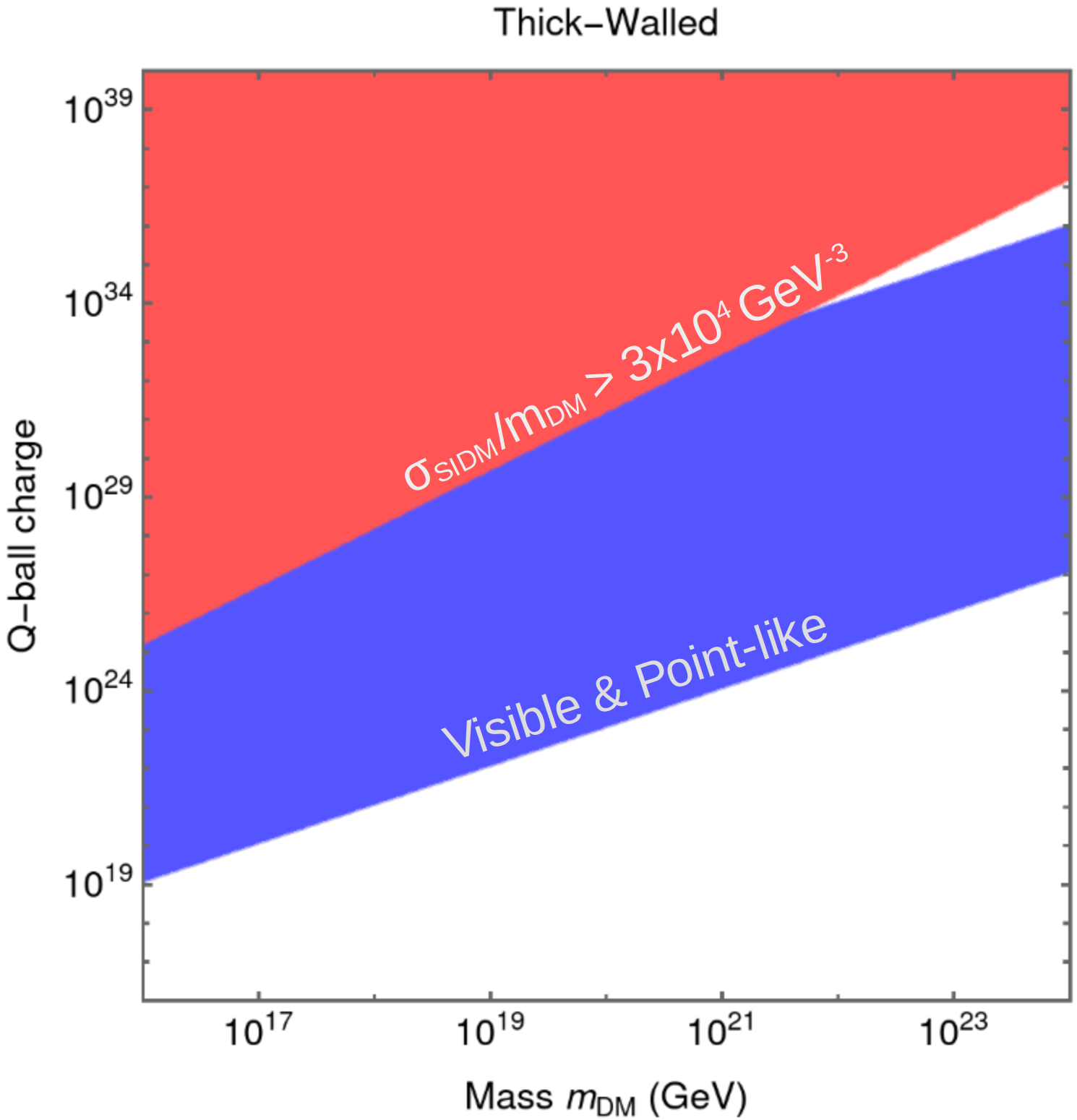}
\caption{The excluded and visible parameter space for two simplified Q-ball models. The blue shaded regions are the visible and point-like regions where $\lambda_{DM}<1200\;\text{GeV}$ and $R>1\;\text{mm}$. The red shaded regions cover model space where the SIDM bounds are exceeded, i.e. $\sigma_{SIDM}/M > 1 \;\text{cm}^2/\text{g}$ ~\cite{Enqvist:2001jd,Gresham:2018anj,Robertson:2016xjh}. The top frame shows the simplified thin-walled Q-ball model space while the bottom frame shows the simplified thick-walled Q-ball model space.}
\label{fig:ssimpleqballs}
\end{figure}

Fig.~\ref{fig:ssimpleqballs} shows the regions of parameter space which are already excluded and those which would be visible to an ideal opto-mechanical array. We find that such a search would be sensitive to a significant portion of the parameter space, probing several orders of magnitude lower in charge than generic SIDM bounds~\cite{Enqvist:2001jd}. 

Since the thin-walled Q-ball radius scales with $Q^{1/3}$ whereas the thick-walled Q-ball radius scales as $Q^{1/4}$, the thin-walled Q-balls grow larger for a given charge. This means that for a given mass, the geometric cross-section breaches the SIDM bounds at lower charges for the thin-walled case. Similarly, thin-walled Q-balls exceed the point-like maximum radius at lower charge. This ultimately means, that there is a significantly smaller visible and point-like region of parameter space for thin-walled Q-balls.

While opto-mechanical arrays are not effective in the detection of thin-walled Q-balls, there is a sizable region of parameter space in which thick-walled Q-balls are visible to a detector like the one described in Sec.~\ref{sec:sec2}.

\subsubsection{Thin-walled Q-balls}
By setting the vacuum expectation value in of the Q-ball field close to $\varphi_0$, the Q-ball model approaches the thin wall approximation \cite{cotner_kusenko_2016}. Note that $\varphi_0$ is a model-specific parameter proportional to the $\varphi$ discussed in the preceding subsection. In this regime, the radial profile of the Q-ball is essentially step-like, i.e. an infinitely thin transition ``wall". The governing equation for this Q-ball type can then be expressed as

\begin{align}
    \frac{Q}{M} \equiv const = \frac{8}{3} \pi \varphi_0^2 Q^{-1} R^3,
    \label{eq:qballRadius_TypeI}
\end{align}
where Q is the charge, M is the Q-ball mass, and R is the radius of the Q-ball. This shows that the mass grows linearly with the charge for a given choice of model parameters. Setting the maximum radius for the Q-ball equal to the scale of the proposed sensor spacing (i.e. 1mm), we find that the region for the Q-ball parameters which is visible to a detector such as that discussed in Sec.~\ref{sec:sec2} contains the Planck mass. As expected, the charge in this region of parameter space at the Planck scale is relatively large, but possible to accommodate by choice of model.

\begin{figure}
    \includegraphics[width=3.5in]{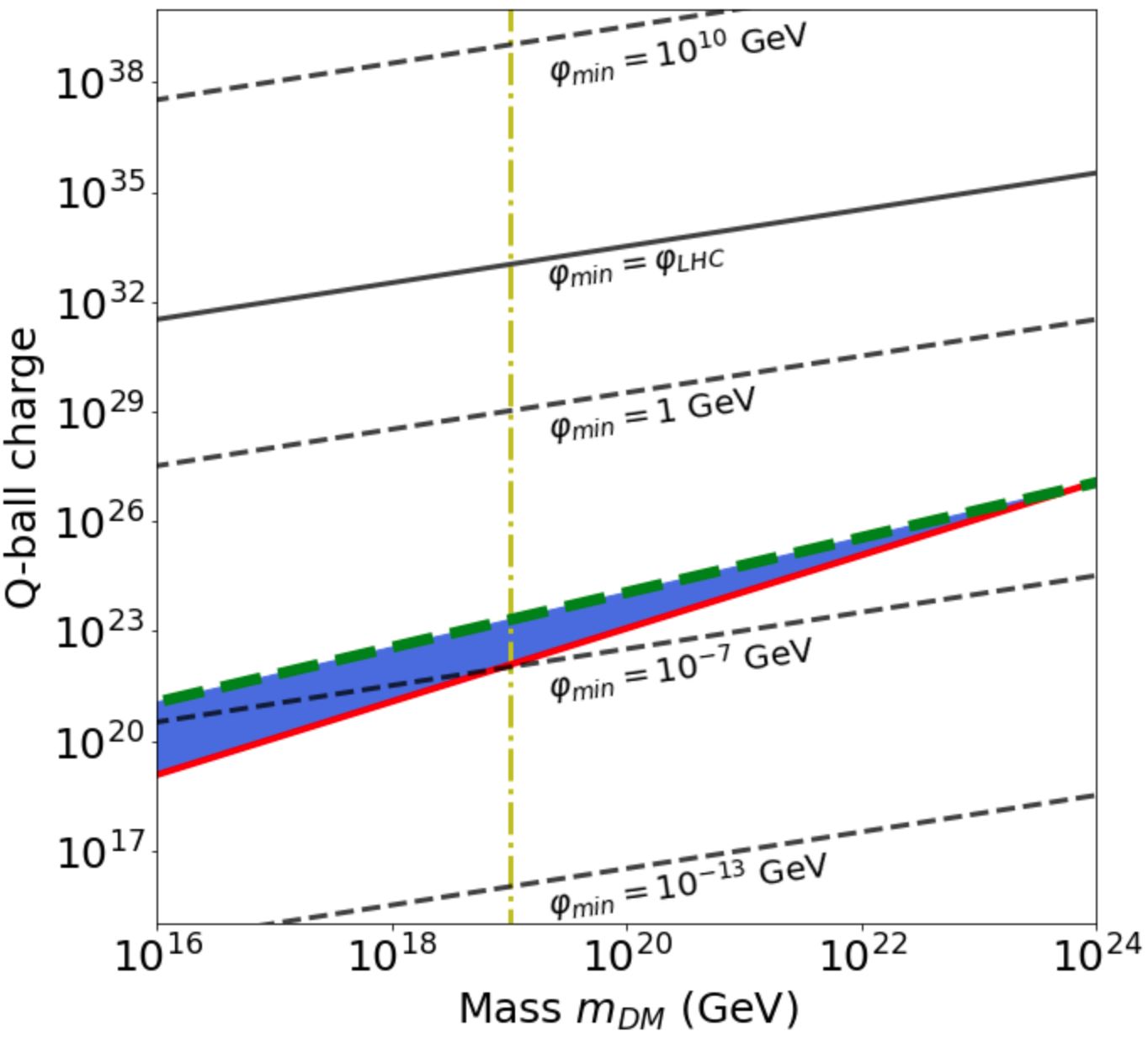}
    \caption{Here we show the parameter space of thin-walled Q-balls. The blue shaded region above the red line and below the dashed green line is where the visible point-like parameter space lies. The dashed black lines are contours of constant $\varphi_{min}$, which is the model-dependent mass scale. These values are computed for a Q-ball radius of 1~mm. The solid black line is a contour along $\varphi_{min} = 10~\mathrm{TeV}$. The red line represents the minimum charge-to-mass ratio visible to a sensor array like that discussed in Sec.~\ref{sec:sec2} and the green line is the contour where the Q-ball radius is 1~mm, i.e. the largest point-like radius. The yellow line is at the Planck mass. }
    
    \label{fig:Type1}
\end{figure}

Fig.~\ref{fig:Type1} shows the parameter space of thin-walled Q-balls. We find that there exists a region of parameter space which is both visible and point-like to a detector like the one discussed in Sec.~\ref{sec:sec2} for charges between about $10^{20}$ and  $10^{23}$. In the figure, $\varphi_{min}$, the minimum value for the model dependent parameter $\varphi_0$, is obtained by setting the Q-ball radius to 1 mm. Note that the visible point-like parameter space below the green line and above the red line lies within the region where $\varphi$ might fall below 10 TeV and could potentially produce visible signatures at the LHC.

The final constraint to check is the black hole limit, which can be expressed as
\begin{align}
    R_{sch} < \left (\frac{3 Q^2}{8 \pi M \varphi_0^2} \right )^{1/3}.
    \label{eq:blackholelim_TypeI_gen}
\end{align}
Therefore, the Q-ball's charge must be large enough for a given mass and $\varphi$ in order to prevent the Q-ball from becoming a singularity,
\begin{align}
    Q > \left (\frac{64 \pi \varphi_0^2 G^3 M^4}{3 } \right)^{1/2}.
    \label{eq:blackholelim_TypeI}
\end{align}
Assuming $\varphi_{min}=10\;\mathrm{TeV}$, this condition is met everywhere in the parameter space shown in Fig.~\ref{fig:Type1}.

\subsubsection{Thick-walled Q-balls with flat potential}
Thick-walled Q-balls are gauge mediated field configurations with flat potentials in which the radial profile decreases monotonically to zero from it's maximum at the Q-ball center\cite{Enqvist:2001jd}. These are fuzzier analogs of thin-walled Q-balls which can be a feature of many BSM models~\cite{Kusenko:1997si,Ponton:2019hux}. In super-symmetric examples of these models, e.g. ref \cite{Kusenko:1997si}, baryonic Q-balls survive until the present day when they confine a baryon charge of 10$^{8}$ - 10$^{20}$. Likewise, lepton balls require some nominal charge in these models to survive until the present day. Thick-walled Q-balls can also be cast into simple Higgs-portal models.

The mass and characteristic radius of thick-walled Q-balls are ~\cite{Enqvist:2001jd},
\begin{align}
    M&= \frac{4 \pi}{3} \sqrt{2} Q^{3/4} \varphi \;\; \textrm{and}\\
    R&= \frac{Q^{1/4}}{\sqrt{2} \varphi},
    \label{eq:qballRadius_TypII}
\end{align}
which can be rearranged to find the charge-to-mass ratio,
 \begin{align}
    \frac{Q}{M}=\frac{3}{4\pi\sqrt{2} \varphi} Q^{1/4}.
    \label{eq:qballQtoM}
\end{align}

\begin{figure}
    \includegraphics[width=3.5in]{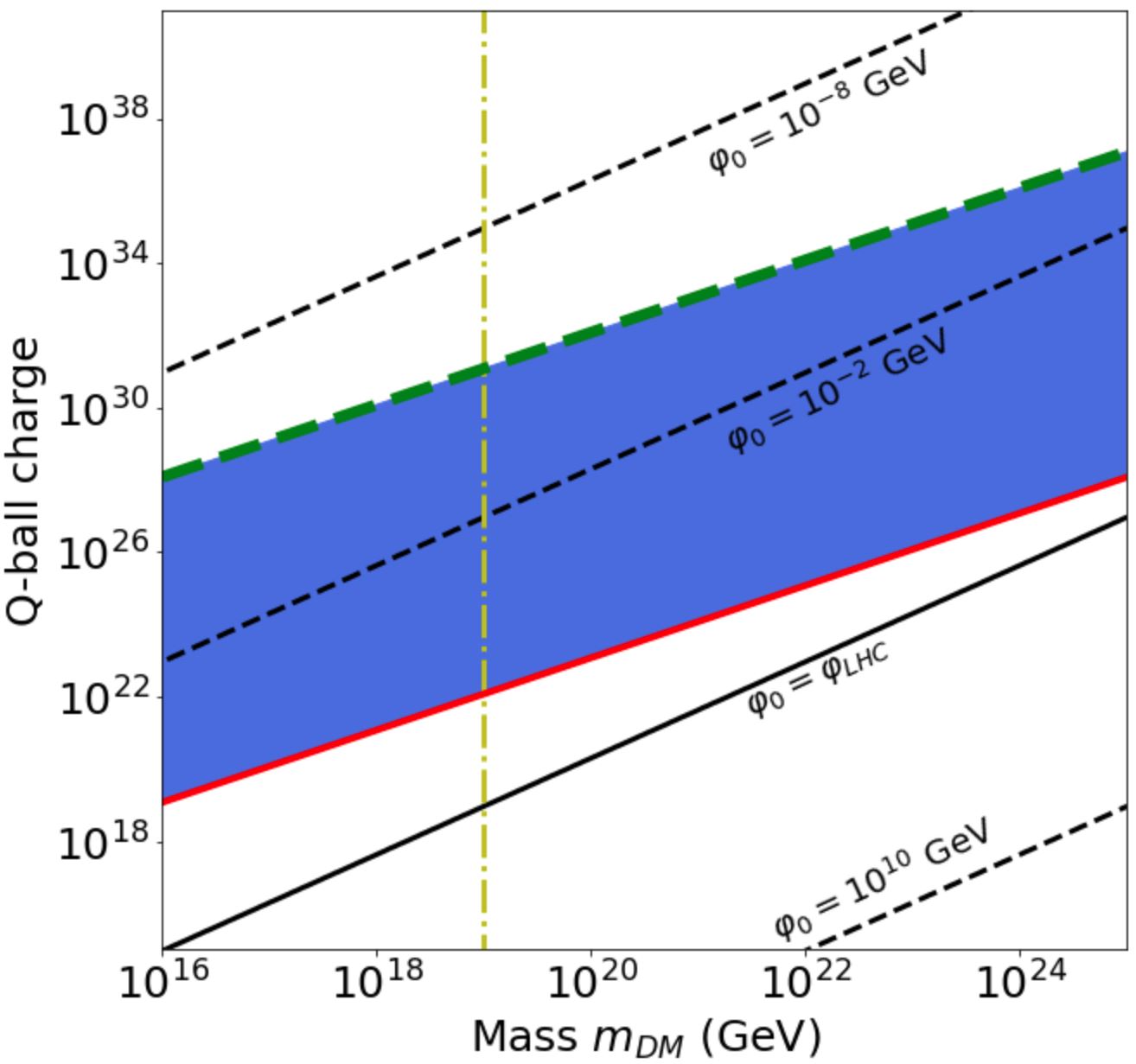}
    \caption{Same as Fig.~\ref{fig:Type1} but for thick-walled Q-balls with flat potentials. The black dashed lines are contours of constant $\varphi_0$, which is the model-dependent mass scale. The blue shaded region between the (dashed) green and red lines is the visible point-like parameter space (where both lines are the same as in Fig.~\ref{fig:Type1}). The yellow line is the Planck mass.}
    \label{fig:Type2}
\end{figure}

Fig.~\ref{fig:Type2} shows the analogous parameter space for thick-walled Q-balls. We find that within the visible point-like region around the Planck scale there is a broad range of charges that are accessible to a detector like the one discussed in Sec.~\ref{sec:sec2}. We find that there exists a region of parameter space which is both visible and point-like to our detector for charges between about $10^{20}$ and  $10^{30}$ and masses between about $10^{16}\;\textrm{GeV}$ and  $10^{26}\;\textrm{GeV}$.

Next, we compute the parameters which push the Q-ball into forming a black hole. For this, it is helpful to rearrange the expressions in Eq.~\ref{eq:qballRadius_TypII} into a model-independent expression for the charge-to-mass ratio,
\begin{align}
   \frac{Q}{M}  = \frac{3}{4 \pi } R.
    \label{eq:gen_Q_M}
\end{align}
Using the equations above and $R_{sch}$, we can then determine the limit on the Q-ball radius and its charge-to-mass ratio. The two equivalent bounds are given by,
\begin{align}
    R > \left( \frac{8\pi G Q}{3  } \right)^{1/2}\;\; \textrm{and} \;\; Q > \frac{3GM^2}{2\pi }.
    \label{eq:blackholelim_TypeII}
\end{align}
This condition is met everywhere in the parameter space shown in Fig.~\ref{fig:Type2}.

\begin{figure}
    \includegraphics[width=3.5in]{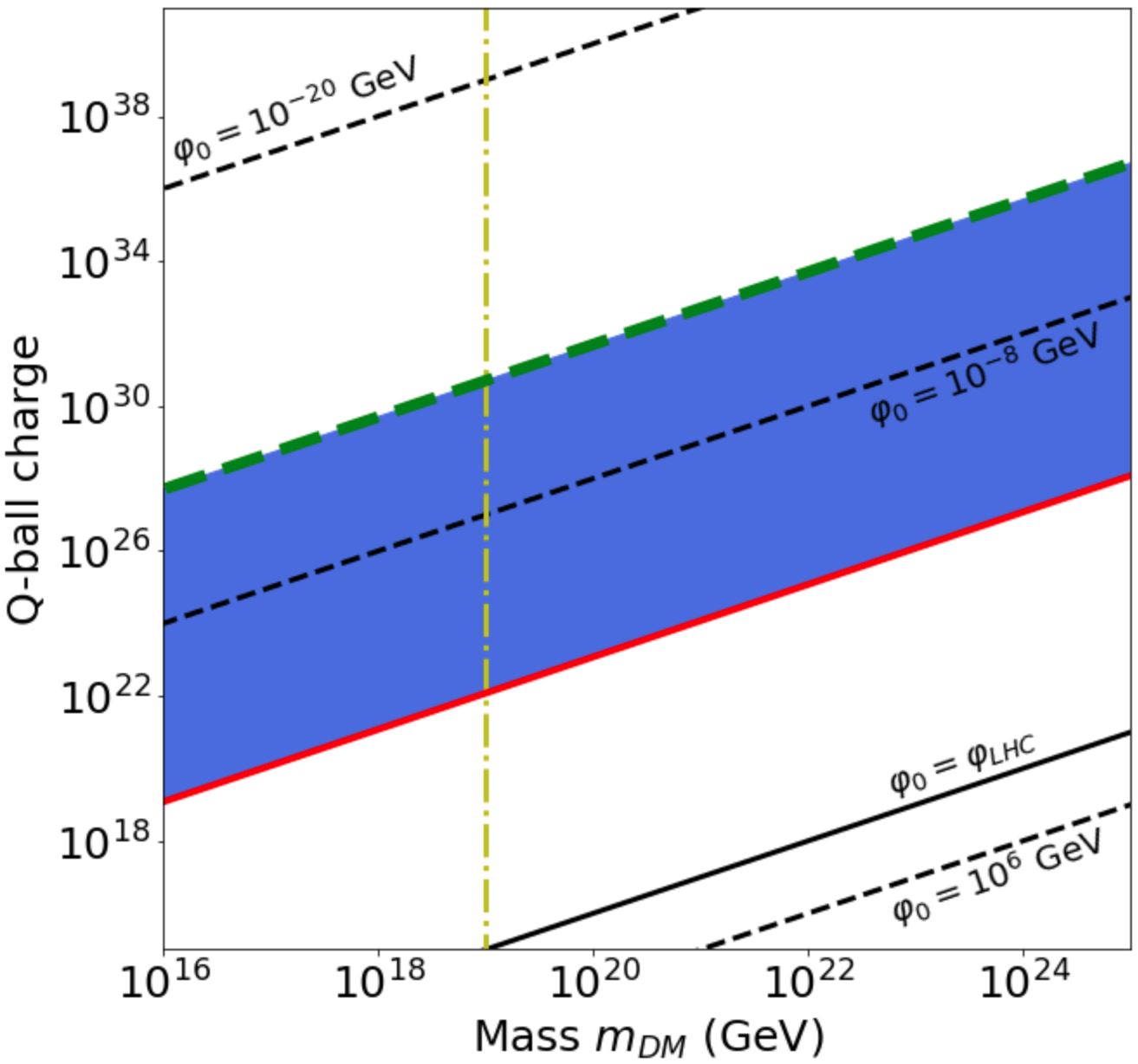}
    \caption{As in Fig.~\ref{fig:Type2} but for thick-walled Q-balls with logarithmic potentials.}
    \label{fig:Type3}
\end{figure}
\subsubsection{Thick-walled Q-balls with logarithmic potential}

Like the thick-walled Q-balls previously discussed, these Q-balls are thick-walled ``fuzzy" field configurations. However, these Q-balls have logarithmic potentials and are exemplified by gravitationally mediated models~\cite{Enqvist:2001jd}. The salient feature of this type of Q-ball is the model dependent constant radius. That is, given a small-scale mass (which is proportional to scalar field constants in the model), we get the following constant radius,
\begin{align}
    R \equiv \kappa \varphi_{0}^{-1} = \kappa \frac{Q}{M},
    \label{eq:qballRadius_TypeIII}
\end{align}
where  $\kappa$ is a positive model-dependent parameter~\cite{Enqvist:2001jd}.

Fig.~\ref{fig:Type3} shows the parameter space of thick-walled Q-balls with a logarithmic potential. We find that there exists a region of parameter space which is both visible and point-like to our detector for charges between about $10^{20}$ and  $10^{30}$. However, the range of the $\varphi_{0}$ parameter expected in the visible point-like region of both thick-walled models discussed here is exceptionally small compared to the weak-scale. Therefore, these thick-walled Q-ball models may also produce visible signatures at the LHC.

The black hole limit for thick-walled Q-balls with logarithmic potentials is,
\begin{align}
    R_{sch} = \frac{2 G \kappa Q}{ R},
    \label{eq:blackholelim_TypeIII_gen}
\end{align}
and the resulting bound on Q is,
\begin{align}
    Q > \frac{2GM^2}{\kappa}.
    \label{eq:blackholelim_TypeIII}
\end{align}
Therefore, the model-dependent parameter $\kappa$ must be sufficiently small for large Q. We find that this condition is met everywhere in the parameter space shown in Fig.~\ref{fig:Type3} for $10^{-2}\lesssim \kappa \lesssim 10^2$.

\subsection{Superheavy DM Candidates}
DM candidates that interact purely gravitationally can be produced in many ways. Here we considered two general production mechanisms. In the first case, particle dark matter up to the Planck scale is produced through its coupling to gravity. In the second case, the DM candidate is the extremal relic left behind after the evaporation of a gravitationally collapsed object such as a primordial black hole. 

We note that for primordial black holes composing most of the dark matter, the literature focuses on the parameter space where the mass of the PBH is above a bound set by cosmological and galactic observations. In particular, the minimum PBH dark matter mass which is set by the CMB signatures of the evaporation of these candidates is around 10$^{16}$ kg ~\cite{Arbey:2019vqx, Villanueva-Domingo:2021spv}. Since this lower bound is much larger than the Planck scale, we consider the extremal primordial black hole relics left behind after PBH evaporation.

\begin{figure}[ht]

    \includegraphics[width=1\linewidth]{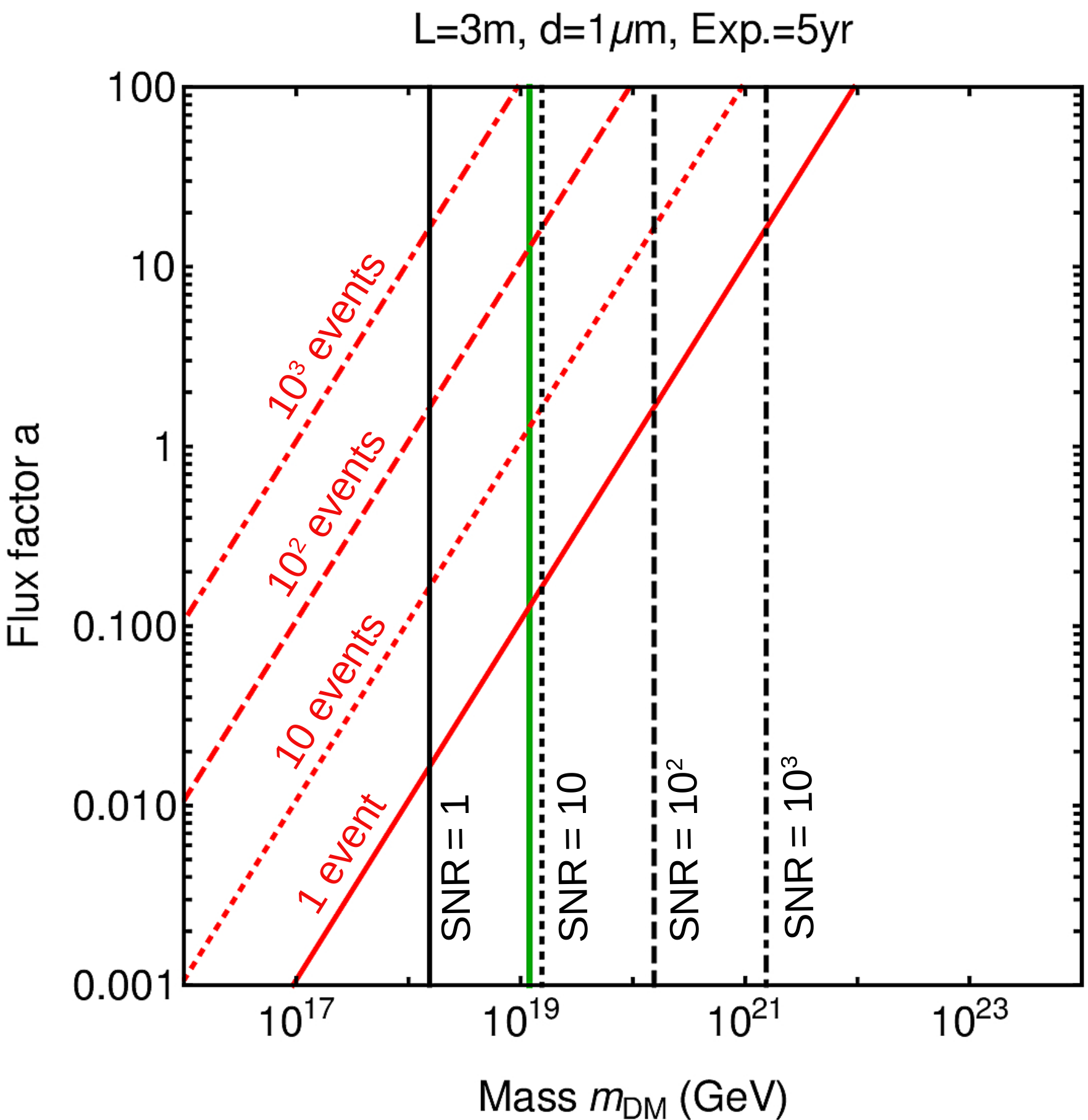}
    \caption{The black lines are contours of constant SNR and the red lines are contours of constant event rate. The green line is the Planck mass. These rates and signal-to-noise ratios are calculated for an array of opto-mechanical sensors like those discussed in Sec.~\ref{sec:sec2} with a  side length $L=3\;\text{m}$ and sensor spacing $d = 1\;\mu\text{m}$ for an exposure of 5 years.}
    \label{fig: pbhGravDM}
\end{figure}

Fig.~\ref{fig: pbhGravDM} shows the rates and signal-to-noise ratios that could be expected for gravitationally interacting Planck-scale DM for an idealized opto-mechanical array of sensors like those discussed in Sec.\ref{sec:sec2}. We find that since the SNR increases with DM mass and the rate decreases, one must maximize the side length and minimize the sensor spacing in order to explore the mass range around the Planck scale. We note that the finite material density drives the sensor mass down as you decrease the sensor separation since the size of the accelerometer also needs to be smaller. Additionally, even mild local overdensities can significantly improve the mass reach of these setups. We note that such over densities can be the result of astrophysical mechanisms such as local streams or gravitational focusing. While the details of such mechanisms are beyond the scope of this study, the density of DM in the solar system is subject to very large uncertainties and values of $a$ up to $10^3$ are not ruled out~\cite{Read:2014qva,Pitjev:2013sfa}. 

\subsubsection{Gravitationally produced particle DM}
WIMPZILLAs are superheavy candidates for dark matter that are the result of the gravitational production of particles during or immediately after the inflationary epoch \cite{chung_crotty_kolb_riotto_2001}. For example, for nonthermal production mechanisms \cite{Chung:1998ua}, which presume a nonequilibrium state at freeze out, a sufficient condition to obtain a DM candidate is that the annihilation rate be smaller than the expansion rate. Furthermore, WIMPZILLA candidates are not fundamentally constrained to a particular charge range~\cite{Chung:1998zb}. The WIMPZILLA charge is completely model-dependent, although the phenomenologically large charges required for visibility would be increasingly difficult to justify from a fundamental particle point of view. While the proposed production mechanisms depend on model-dependent details of the thermal history of the universe, primarily the reheating temperature to arrive at the correct relic abundance, it should be noted that producing WIMPZILLAs within one or two orders of magnitude of the Planck mass is increasingly difficult. 

Gravitational production of ultraheavy DM up to the Planck mass has also been proposed through the evaporation of black holes after an epoch of black hole domination in the early universe~\cite{Hooper:2019gtx}. This relatively simple scenario is one in which a population of black holes that comes to dominate the early universe populates the dark sector via Hawking radiation. Since gravity couples universally to mass, DM and dark radiation can be produced readily when the temperature of the black hole is above the mass of the dark particle and can therefore generate DM up to the Planck scale.

\subsubsection{Extremal relics of evaporating singularities}

Models of Planck-scale gravitational relics can follow from the assumption that the evaporation process of Hawking Radiation halts near the Planck scale, where quantum gravity effects become important. This could leave behind a DM candidate in the form of a black hole relic \cite{Bernal:2020bjf, lehmann_johnson_profumo_schwemberger_2019,Bai:2019zcd,MacGibbon:1987my} or the Planck-scale relic of the black hole-like endpoint of gravitational collapse~\cite{Salvio:2019llz,Aydemir:2020xfd}. 

A the result of Hawking radiation that evaporates a singularity containing a conserved charge. Generically, the equilibrium-seeking evaporation process dynamically drives this charge to be of order 1. Specifically, the preferential Hawking radiation of particles with charge like that of the collapsed object drives the charge of the relic to be order-1. Although the extremal relic contains a conserved charge, the magnitude is tiny. Therefore, we focus on the gravitational interaction of the relic and subject these to the discussion of the gravitational limit in Sec.~\ref{sec:sec2}.

While exclusively gravitationally interacting PBH  relics were previously discussed as undetectable \cite{lehmann_johnson_profumo_schwemberger_2019, Carr:2021bzv}, we propose that the Windchime experiment would be ideally suited to detect and identify these DM candidates. Notably, the mass of these relics should range from the Planck-scale and up, making them particularly well motivated super-Planckian DM candidates for Windchime.

\section{Conclusion}
\label{sec:sec4}
There is a variety of unexplored DM models within the sensitivity bounds of an opto-mechanical experiment such as Windchime. Due to their experimental design, opto-mechanical arrays must look for tracks left by point-like heavy particles. By choosing the sensor mass and separation, one can improve the probability of obtaining a significant event rate during a given exposure. We find that Planck-scale DM coupling in the Coulombic regime through an unscreened new force and Planck-scale DM coupling purely gravitationally both leave similar visible signatures in these detectors. We have shown that there exist significant regions of parameter space which are visible to experiments like Windchime and are well-motivated while remaining reasonably model-independent.

If the DM interacts via a new long-range force saturating current fifth-force constraints~\cite{Schlamminger:2007ht,Wagner:2012ui,Abbott:2021npy}, the sensitivity lies in a region of large charge-to-mass ratio, $\lambda_{DM} \geq 1200$ GeV$^{-1}$, which can be realized in the context of composite DM or non-topological soliton models. These classes of models can efficiently confine large charges into tiny volumes without necessitating huge fundamental charges. There exist other models of compact composite objects, such as axion stars, that might be visible to opto-mechanical arrays. However, these objects are not point-like at the scale of the detector and so we leave such discussions for future work.

Composite models that confine a charge greater than about $10^{25}$ and have a mass per constituent of less than around $1\;\text{keV}$ would be visible to the opto-mechanical arrays discussed here. Such models are safe from cosmological constraints if they confine that charge before BBN. We conclude that while Windchime-like detectors have a comparatively smaller sensitivity window to thin-walled Q-balls, thick-walled Q-ball models may be visible for charges between $10^{20}$ to $10^{30}$ and for masses masses above $10^{16}\;\text{GeV}$.

When the DM only interacts gravitationally with the detector's sensors, the limiting factors become the event rate and the signal-to-noise ratio. Generally, non-fundamental candidates such as extremal gravitational relics motivate searching for DM above the Planck mass. WIMPZILLAS and other gravitationally produced particles occupy the parameter space below that. While we have not investigated all possible gravitationally coupled Planck-scale candidates, the general conclusions we have drawn in Sec.~\ref{sec:sec3} D. are expected to apply to any interacting particle that couples to the SM only through gravity. We also note that sensitive impulse-based searches such as Windchime remain the only proposed method for directly detecting such Planck-scale gravitationally interacting DM candidates.

\section{Acknowledgements}
\label{sec:sec5}
We thank Dan Carney for helpful comments and fruitful discussions. The work of CB was supported in part by NASA through the NASA Hubble Fellowship Program grant HST-HF2-51451.001-A awarded by the Space Telescope Science Institute, which is operated by the Association of Universities for Research in Astronomy, Inc., for NASA, under contract NAS5-26555 as well as by the European Research Council under grant 742104. CB, BE and RFL are supported by the U.S. Department of Energy, Office of Science, National Quantum Information Science Research Centers, Quantum Science Center.

\bibliography{main}

\providecommand{\noopsort}[1]{}\providecommand{\singleletter}[1]{#1}%
\begin{thebibliography}{96}%
\makeatletter
\providecommand \@ifxundefined [1]{%
 \@ifx{#1\undefined}
}%
\providecommand \@ifnum [1]{%
 \ifnum #1\expandafter \@firstoftwo
 \else \expandafter \@secondoftwo
 \fi
}%
\providecommand \@ifx [1]{%
 \ifx #1\expandafter \@firstoftwo
 \else \expandafter \@secondoftwo
 \fi
}%
\providecommand \natexlab [1]{#1}%
\providecommand \enquote  [1]{``#1''}%
\providecommand \bibnamefont  [1]{#1}%
\providecommand \bibfnamefont [1]{#1}%
\providecommand \citenamefont [1]{#1}%
\providecommand \href@noop [0]{\@secondoftwo}%
\providecommand \href [0]{\begingroup \@sanitize@url \@href}%
\providecommand \@href[1]{\@@startlink{#1}\@@href}%
\providecommand \@@href[1]{\endgroup#1\@@endlink}%
\providecommand \@sanitize@url [0]{\catcode `\\12\catcode `\$12\catcode
  `\&12\catcode `\#12\catcode `\^12\catcode `\_12\catcode `\%12\relax}%
\providecommand \@@startlink[1]{}%
\providecommand \@@endlink[0]{}%
\providecommand \url  [0]{\begingroup\@sanitize@url \@url }%
\providecommand \@url [1]{\endgroup\@href {#1}{\urlprefix }}%
\providecommand \urlprefix  [0]{URL }%
\providecommand \Eprint [0]{\href }%
\providecommand \doibase [0]{https://doi.org/}%
\providecommand \selectlanguage [0]{\@gobble}%
\providecommand \bibinfo  [0]{\@secondoftwo}%
\providecommand \bibfield  [0]{\@secondoftwo}%
\providecommand \translation [1]{[#1]}%
\providecommand \BibitemOpen [0]{}%
\providecommand \bibitemStop [0]{}%
\providecommand \bibitemNoStop [0]{.\EOS\space}%
\providecommand \EOS [0]{\spacefactor3000\relax}%
\providecommand \BibitemShut  [1]{\csname bibitem#1\endcsname}%
\let\auto@bib@innerbib\@empty
\bibitem [{\citenamefont {Carney}\ \emph {et~al.}(2021)\citenamefont {Carney}
  \emph {et~al.}}]{Carney:2020xol}%
  \BibitemOpen
  \bibfield  {author} {\bibinfo {author} {\bibfnamefont {D.}~\bibnamefont
  {Carney}} \emph {et~al.},\ }\bibfield  {title} {\bibinfo {title} {{Mechanical
  Quantum Sensing in the Search for Dark Matter}},\ }\href
  {https://doi.org/10.1088/2058-9565/abcfcd} {\bibfield  {journal} {\bibinfo
  {journal} {Quantum Sci. Technol.}\ }\textbf {\bibinfo {volume} {6}},\
  \bibinfo {pages} {024002} (\bibinfo {year} {2021})},\ \Eprint
  {https://arxiv.org/abs/2008.06074} {arXiv:2008.06074 [physics.ins-det]}
  \BibitemShut {NoStop}%
\bibitem [{\citenamefont {Carney}\ \emph {et~al.}(2020)\citenamefont {Carney},
  \citenamefont {Ghosh}, \citenamefont {Krnjaic},\ and\ \citenamefont
  {Taylor}}]{Carney:2019pza}%
  \BibitemOpen
  \bibfield  {author} {\bibinfo {author} {\bibfnamefont {D.}~\bibnamefont
  {Carney}}, \bibinfo {author} {\bibfnamefont {S.}~\bibnamefont {Ghosh}},
  \bibinfo {author} {\bibfnamefont {G.}~\bibnamefont {Krnjaic}},\ and\ \bibinfo
  {author} {\bibfnamefont {J.~M.}\ \bibnamefont {Taylor}},\ }\bibfield  {title}
  {\bibinfo {title} {{Proposal for gravitational direct detection of dark
  matter}},\ }\href {https://doi.org/10.1103/PhysRevD.102.072003} {\bibfield
  {journal} {\bibinfo  {journal} {Phys. Rev. D}\ }\textbf {\bibinfo {volume}
  {102}},\ \bibinfo {pages} {072003} (\bibinfo {year} {2020})},\ \Eprint
  {https://arxiv.org/abs/1903.00492} {arXiv:1903.00492 [hep-ph]} \BibitemShut
  {NoStop}%
\bibitem [{\citenamefont {Moore}\ and\ \citenamefont
  {Geraci}(2021)}]{Moore:2020awi}%
  \BibitemOpen
  \bibfield  {author} {\bibinfo {author} {\bibfnamefont {D.~C.}\ \bibnamefont
  {Moore}}\ and\ \bibinfo {author} {\bibfnamefont {A.~A.}\ \bibnamefont
  {Geraci}},\ }\bibfield  {title} {\bibinfo {title} {{Searching for new physics
  using optically levitated sensors}},\ }\href
  {https://doi.org/10.1088/2058-9565/abcf8a} {\bibfield  {journal} {\bibinfo
  {journal} {Quantum Sci. Technol.}\ }\textbf {\bibinfo {volume} {6}},\
  \bibinfo {pages} {014008} (\bibinfo {year} {2021})},\ \Eprint
  {https://arxiv.org/abs/2008.13197} {arXiv:2008.13197 [quant-ph]} \BibitemShut
  {NoStop}%
\bibitem [{\citenamefont {Monteiro}\ \emph {et~al.}(2020)\citenamefont
  {Monteiro}, \citenamefont {Afek}, \citenamefont {Carney}, \citenamefont
  {Krnjaic}, \citenamefont {Wang},\ and\ \citenamefont
  {Moore}}]{Monteiro:2020wcb}%
  \BibitemOpen
  \bibfield  {author} {\bibinfo {author} {\bibfnamefont {F.}~\bibnamefont
  {Monteiro}}, \bibinfo {author} {\bibfnamefont {G.}~\bibnamefont {Afek}},
  \bibinfo {author} {\bibfnamefont {D.}~\bibnamefont {Carney}}, \bibinfo
  {author} {\bibfnamefont {G.}~\bibnamefont {Krnjaic}}, \bibinfo {author}
  {\bibfnamefont {J.}~\bibnamefont {Wang}},\ and\ \bibinfo {author}
  {\bibfnamefont {D.~C.}\ \bibnamefont {Moore}},\ }\bibfield  {title} {\bibinfo
  {title} {{Search for composite dark matter with optically levitated
  sensors}},\ }\href {https://doi.org/10.1103/PhysRevLett.125.181102}
  {\bibfield  {journal} {\bibinfo  {journal} {Phys. Rev. Lett.}\ }\textbf
  {\bibinfo {volume} {125}},\ \bibinfo {pages} {181102} (\bibinfo {year}
  {2020})},\ \Eprint {https://arxiv.org/abs/2007.12067} {arXiv:2007.12067
  [hep-ex]} \BibitemShut {NoStop}%
\bibitem [{\citenamefont {Ghosh}\ \emph {et~al.}(2020)\citenamefont {Ghosh},
  \citenamefont {Carney}, \citenamefont {Shawhan},\ and\ \citenamefont
  {Taylor}}]{Ghosh:2019rsc}%
  \BibitemOpen
  \bibfield  {author} {\bibinfo {author} {\bibfnamefont {S.}~\bibnamefont
  {Ghosh}}, \bibinfo {author} {\bibfnamefont {D.}~\bibnamefont {Carney}},
  \bibinfo {author} {\bibfnamefont {P.}~\bibnamefont {Shawhan}},\ and\ \bibinfo
  {author} {\bibfnamefont {J.~M.}\ \bibnamefont {Taylor}},\ }\bibfield  {title}
  {\bibinfo {title} {{Backaction-evading impulse measurement with mechanical
  quantum sensors}},\ }\href {https://doi.org/10.1103/PhysRevA.102.023525}
  {\bibfield  {journal} {\bibinfo  {journal} {Phys. Rev. A}\ }\textbf {\bibinfo
  {volume} {102}},\ \bibinfo {pages} {023525} (\bibinfo {year} {2020})},\
  \Eprint {https://arxiv.org/abs/1910.11892} {arXiv:1910.11892 [quant-ph]}
  \BibitemShut {NoStop}%
\bibitem [{\citenamefont {Kawasaki}(2019)}]{Kawasaki:2018xak}%
  \BibitemOpen
  \bibfield  {author} {\bibinfo {author} {\bibfnamefont {A.}~\bibnamefont
  {Kawasaki}},\ }\bibfield  {title} {\bibinfo {title} {{Search for
  Kilogram-scale Dark Matter with Precision Displacement Sensors}},\ }\href
  {https://doi.org/10.1103/PhysRevD.99.023005} {\bibfield  {journal} {\bibinfo
  {journal} {Phys. Rev. D}\ }\textbf {\bibinfo {volume} {99}},\ \bibinfo
  {pages} {023005} (\bibinfo {year} {2019})},\ \Eprint
  {https://arxiv.org/abs/1809.00968} {arXiv:1809.00968 [physics.ins-det]}
  \BibitemShut {NoStop}%
\bibitem [{\citenamefont {Hall}\ \emph {et~al.}(2018)\citenamefont {Hall},
  \citenamefont {Adhikari}, \citenamefont {Frolov}, \citenamefont {M\"uller},
  \citenamefont {Pospelov},\ and\ \citenamefont {Adhikari}}]{Hall:2016usm}%
  \BibitemOpen
  \bibfield  {author} {\bibinfo {author} {\bibfnamefont {E.~D.}\ \bibnamefont
  {Hall}}, \bibinfo {author} {\bibfnamefont {R.~X.}\ \bibnamefont {Adhikari}},
  \bibinfo {author} {\bibfnamefont {V.~V.}\ \bibnamefont {Frolov}}, \bibinfo
  {author} {\bibfnamefont {H.}~\bibnamefont {M\"uller}}, \bibinfo {author}
  {\bibfnamefont {M.}~\bibnamefont {Pospelov}},\ and\ \bibinfo {author}
  {\bibfnamefont {R.~X.}\ \bibnamefont {Adhikari}},\ }\bibfield  {title}
  {\bibinfo {title} {{Laser Interferometers as Dark Matter Detectors}},\ }\href
  {https://doi.org/10.1103/PhysRevD.98.083019} {\bibfield  {journal} {\bibinfo
  {journal} {Phys. Rev. D}\ }\textbf {\bibinfo {volume} {98}},\ \bibinfo
  {pages} {083019} (\bibinfo {year} {2018})},\ \Eprint
  {https://arxiv.org/abs/1605.01103} {arXiv:1605.01103 [gr-qc]} \BibitemShut
  {NoStop}%
\bibitem [{\citenamefont {Manley}\ \emph {et~al.}(2020)\citenamefont {Manley},
  \citenamefont {Wilson}, \citenamefont {Stump}, \citenamefont {Grin},\ and\
  \citenamefont {Singh}}]{Manley:2019vxy}%
  \BibitemOpen
  \bibfield  {author} {\bibinfo {author} {\bibfnamefont {J.}~\bibnamefont
  {Manley}}, \bibinfo {author} {\bibfnamefont {D.}~\bibnamefont {Wilson}},
  \bibinfo {author} {\bibfnamefont {R.}~\bibnamefont {Stump}}, \bibinfo
  {author} {\bibfnamefont {D.}~\bibnamefont {Grin}},\ and\ \bibinfo {author}
  {\bibfnamefont {S.}~\bibnamefont {Singh}},\ }\bibfield  {title} {\bibinfo
  {title} {{Searching for Scalar Dark Matter with Compact Mechanical
  Resonators}},\ }\href {https://doi.org/10.1103/PhysRevLett.124.151301}
  {\bibfield  {journal} {\bibinfo  {journal} {Phys. Rev. Lett.}\ }\textbf
  {\bibinfo {volume} {124}},\ \bibinfo {pages} {151301} (\bibinfo {year}
  {2020})},\ \Eprint {https://arxiv.org/abs/1910.07574} {arXiv:1910.07574
  [astro-ph.IM]} \BibitemShut {NoStop}%
\bibitem [{\citenamefont {Manley}\ \emph {et~al.}(2021)\citenamefont {Manley},
  \citenamefont {Chowdhury}, \citenamefont {Grin}, \citenamefont {Singh},\ and\
  \citenamefont {Wilson}}]{Manley:2020mjq}%
  \BibitemOpen
  \bibfield  {author} {\bibinfo {author} {\bibfnamefont {J.}~\bibnamefont
  {Manley}}, \bibinfo {author} {\bibfnamefont {M.~D.}\ \bibnamefont
  {Chowdhury}}, \bibinfo {author} {\bibfnamefont {D.}~\bibnamefont {Grin}},
  \bibinfo {author} {\bibfnamefont {S.}~\bibnamefont {Singh}},\ and\ \bibinfo
  {author} {\bibfnamefont {D.~J.}\ \bibnamefont {Wilson}},\ }\bibfield  {title}
  {\bibinfo {title} {{Searching for vector dark matter with an optomechanical
  accelerometer}},\ }\href {https://doi.org/10.1103/PhysRevLett.126.061301}
  {\bibfield  {journal} {\bibinfo  {journal} {Phys. Rev. Lett.}\ }\textbf
  {\bibinfo {volume} {126}},\ \bibinfo {pages} {061301} (\bibinfo {year}
  {2021})},\ \Eprint {https://arxiv.org/abs/2007.04899} {arXiv:2007.04899
  [quant-ph]} \BibitemShut {NoStop}%
\bibitem [{\citenamefont {Lehmann}\ \emph {et~al.}(2019)\citenamefont
  {Lehmann}, \citenamefont {Johnson}, \citenamefont {Profumo},\ and\
  \citenamefont {Schwemberger}}]{lehmann_johnson_profumo_schwemberger_2019}%
  \BibitemOpen
  \bibfield  {author} {\bibinfo {author} {\bibfnamefont {B.~V.}\ \bibnamefont
  {Lehmann}}, \bibinfo {author} {\bibfnamefont {C.}~\bibnamefont {Johnson}},
  \bibinfo {author} {\bibfnamefont {S.}~\bibnamefont {Profumo}},\ and\ \bibinfo
  {author} {\bibfnamefont {T.}~\bibnamefont {Schwemberger}},\ }\bibfield
  {title} {\bibinfo {title} {Direct detection of primordial black hole relics
  as dark matter},\ }\href {https://doi.org/10.1088/1475-7516/2019/10/046}
  {\bibfield  {journal} {\bibinfo  {journal} {Journal of Cosmology and
  Astroparticle Physics}\ }\textbf {\bibinfo {volume} {2019}}\bibinfo  {number}
  { (10)},\ \bibinfo {pages} {046–046}}\BibitemShut {NoStop}%
\bibitem [{\citenamefont {Meissner}\ and\ \citenamefont
  {Nicolai}(2019)}]{Meissner:2018cay}%
  \BibitemOpen
\bibfield  {number} {  }\bibfield  {author} {\bibinfo {author} {\bibfnamefont
  {K.~A.}\ \bibnamefont {Meissner}}\ and\ \bibinfo {author} {\bibfnamefont
  {H.}~\bibnamefont {Nicolai}},\ }\bibfield  {title} {\bibinfo {title} {{Planck
  Mass Charged Gravitino Dark Matter}},\ }\href
  {https://doi.org/10.1103/PhysRevD.100.035001} {\bibfield  {journal} {\bibinfo
   {journal} {Phys. Rev. D}\ }\textbf {\bibinfo {volume} {100}},\ \bibinfo
  {pages} {035001} (\bibinfo {year} {2019})},\ \Eprint
  {https://arxiv.org/abs/1809.01441} {arXiv:1809.01441 [hep-ph]} \BibitemShut
  {NoStop}%
\bibitem [{\citenamefont {Arafune}\ \emph {et~al.}(2000)\citenamefont
  {Arafune}, \citenamefont {Yoshida}, \citenamefont {Nakamura},\ and\
  \citenamefont {Ogure}}]{Arafune:2000yv}%
  \BibitemOpen
  \bibfield  {author} {\bibinfo {author} {\bibfnamefont {J.}~\bibnamefont
  {Arafune}}, \bibinfo {author} {\bibfnamefont {T.}~\bibnamefont {Yoshida}},
  \bibinfo {author} {\bibfnamefont {S.}~\bibnamefont {Nakamura}},\ and\
  \bibinfo {author} {\bibfnamefont {K.}~\bibnamefont {Ogure}},\ }\bibfield
  {title} {\bibinfo {title} {{Experimental bounds on masses and fluxes of
  nontopological solitons}},\ }\href
  {https://doi.org/10.1103/PhysRevD.62.105013} {\bibfield  {journal} {\bibinfo
  {journal} {Phys. Rev. D}\ }\textbf {\bibinfo {volume} {62}},\ \bibinfo
  {pages} {105013} (\bibinfo {year} {2000})},\ \Eprint
  {https://arxiv.org/abs/hep-ph/0005103} {arXiv:hep-ph/0005103} \BibitemShut
  {NoStop}%
\bibitem [{\citenamefont {Piotrowski}\ \emph {et~al.}(2020)\citenamefont
  {Piotrowski}, \citenamefont {Ma\l{}ek}, \citenamefont {Mankiewicz},
  \citenamefont {Soko\l{}owski}, \citenamefont {Wrochna}, \citenamefont
  {Zadro\.zny},\ and\ \citenamefont {\.Zarnecki}}]{Piotrowski:2020ftp}%
  \BibitemOpen
  \bibfield  {author} {\bibinfo {author} {\bibfnamefont {L.~W.}\ \bibnamefont
  {Piotrowski}}, \bibinfo {author} {\bibfnamefont {K.}~\bibnamefont
  {Ma\l{}ek}}, \bibinfo {author} {\bibfnamefont {L.}~\bibnamefont
  {Mankiewicz}}, \bibinfo {author} {\bibfnamefont {M.}~\bibnamefont
  {Soko\l{}owski}}, \bibinfo {author} {\bibfnamefont {G.}~\bibnamefont
  {Wrochna}}, \bibinfo {author} {\bibfnamefont {A.}~\bibnamefont
  {Zadro\.zny}},\ and\ \bibinfo {author} {\bibfnamefont {A.~F.}\ \bibnamefont
  {\.Zarnecki}},\ }\bibfield  {title} {\bibinfo {title} {{Limits on the Flux of
  Nuclearites and Other Heavy Compact Objects from the Pi of the Sky
  Project}},\ }\href {https://doi.org/10.1103/PhysRevLett.125.091101}
  {\bibfield  {journal} {\bibinfo  {journal} {Phys. Rev. Lett.}\ }\textbf
  {\bibinfo {volume} {125}},\ \bibinfo {pages} {091101} (\bibinfo {year}
  {2020})},\ \Eprint {https://arxiv.org/abs/2008.01285} {arXiv:2008.01285
  [astro-ph.HE]} \BibitemShut {NoStop}%
\bibitem [{\citenamefont {Mayotte}(2016)}]{Mayotte:2016ufk}%
  \BibitemOpen
  \bibfield  {author} {\bibinfo {author} {\bibfnamefont {E.~W.}\ \bibnamefont
  {Mayotte}},\ }\emph {\bibinfo {title} {{Searching for slow-developing
  cosmic-ray showers: looking for evidence of exotic primaries at the Pierre
  Auger Observatory}}},\ \href@noop {} {Ph.D. thesis},\ \bibinfo  {school}
  {Colorado School of Mines} (\bibinfo {year} {2016})\BibitemShut {NoStop}%
\bibitem [{\citenamefont {Pshirkov}(2016)}]{Pshirkov:2015cca}%
  \BibitemOpen
  \bibfield  {author} {\bibinfo {author} {\bibfnamefont {M.~S.}\ \bibnamefont
  {Pshirkov}},\ }\bibfield  {title} {\bibinfo {title} {{Prospects for
  strangelet detection with large-scale cosmic ray observatories}},\ }\href
  {https://doi.org/10.1142/S0218271816501030} {\bibfield  {journal} {\bibinfo
  {journal} {Int. J. Mod. Phys. D}\ }\textbf {\bibinfo {volume} {25}},\
  \bibinfo {pages} {1650103} (\bibinfo {year} {2016})},\ \Eprint
  {https://arxiv.org/abs/1509.05553} {arXiv:1509.05553 [astro-ph.IM]}
  \BibitemShut {NoStop}%
\bibitem [{\citenamefont {Cecchini}\ \emph {et~al.}(2008)\citenamefont
  {Cecchini} \emph {et~al.}}]{SLIM:2008bwg}%
  \BibitemOpen
  \bibfield  {author} {\bibinfo {author} {\bibfnamefont {S.}~\bibnamefont
  {Cecchini}} \emph {et~al.} (\bibinfo {collaboration} {SLIM}),\ }\bibfield
  {title} {\bibinfo {title} {{Results of the Search for Strange Quark Matter
  and Q-balls with the SLIM Experiment}},\ }\href
  {https://doi.org/10.1140/epjc/s10052-008-0747-7} {\bibfield  {journal}
  {\bibinfo  {journal} {Eur. Phys. J. C}\ }\textbf {\bibinfo {volume} {57}},\
  \bibinfo {pages} {525} (\bibinfo {year} {2008})},\ \Eprint
  {https://arxiv.org/abs/0805.1797} {arXiv:0805.1797 [hep-ex]} \BibitemShut
  {NoStop}%
\bibitem [{\citenamefont {Ahlers}\ \emph {et~al.}(2018)\citenamefont {Ahlers},
  \citenamefont {Helbing},\ and\ \citenamefont {P\'erez de~los
  Heros}}]{Ahlers:2018mkf}%
  \BibitemOpen
  \bibfield  {author} {\bibinfo {author} {\bibfnamefont {M.}~\bibnamefont
  {Ahlers}}, \bibinfo {author} {\bibfnamefont {K.}~\bibnamefont {Helbing}},\
  and\ \bibinfo {author} {\bibfnamefont {C.}~\bibnamefont {P\'erez de~los
  Heros}},\ }\bibfield  {title} {\bibinfo {title} {{Probing Particle Physics
  with IceCube}},\ }\href {https://doi.org/10.1140/epjc/s10052-018-6369-9}
  {\bibfield  {journal} {\bibinfo  {journal} {Eur. Phys. J. C}\ }\textbf
  {\bibinfo {volume} {78}},\ \bibinfo {pages} {924} (\bibinfo {year} {2018})},\
  \Eprint {https://arxiv.org/abs/1806.05696} {arXiv:1806.05696 [astro-ph.HE]}
  \BibitemShut {NoStop}%
\bibitem [{\citenamefont {Bassan}\ \emph {et~al.}(2016)\citenamefont {Bassan}
  \emph {et~al.}}]{ROG:2015teb}%
  \BibitemOpen
  \bibfield  {author} {\bibinfo {author} {\bibfnamefont {M.}~\bibnamefont
  {Bassan}} \emph {et~al.} (\bibinfo {collaboration} {ROG}),\ }\bibfield
  {title} {\bibinfo {title} {{Dark Matter searches using gravitational wave bar
  detectors: quark nuggets and newtorites}},\ }\href
  {https://doi.org/10.1016/j.astropartphys.2015.12.007} {\bibfield  {journal}
  {\bibinfo  {journal} {Astropart. Phys.}\ }\textbf {\bibinfo {volume} {78}},\
  \bibinfo {pages} {52} (\bibinfo {year} {2016})},\ \Eprint
  {https://arxiv.org/abs/1512.06249} {arXiv:1512.06249 [physics.ins-det]}
  \BibitemShut {NoStop}%
\bibitem [{\citenamefont {Liu}\ and\ \citenamefont
  {Barish}(1988)}]{Liu:1988wn}%
  \BibitemOpen
  \bibfield  {author} {\bibinfo {author} {\bibfnamefont {G.}~\bibnamefont
  {Liu}}\ and\ \bibinfo {author} {\bibfnamefont {B.}~\bibnamefont {Barish}},\
  }\bibfield  {title} {\bibinfo {title} {{Nuclearites Flux Limit From
  Gravitational Wave Detectors}},\ }\href
  {https://doi.org/10.1103/PhysRevLett.61.271} {\bibfield  {journal} {\bibinfo
  {journal} {Phys. Rev. Lett.}\ }\textbf {\bibinfo {volume} {61}},\ \bibinfo
  {pages} {271} (\bibinfo {year} {1988})}\BibitemShut {NoStop}%
\bibitem [{\citenamefont {Price}(1988)}]{Price:1988ge}%
  \BibitemOpen
  \bibfield  {author} {\bibinfo {author} {\bibfnamefont {P.~B.}\ \bibnamefont
  {Price}},\ }\bibfield  {title} {\bibinfo {title} {{Limits on Contribution of
  Cosmic Nuclearites to Galactic Dark Matter}},\ }\href
  {https://doi.org/10.1103/PhysRevD.38.3813} {\bibfield  {journal} {\bibinfo
  {journal} {Phys. Rev. D}\ }\textbf {\bibinfo {volume} {38}},\ \bibinfo
  {pages} {3813} (\bibinfo {year} {1988})}\BibitemShut {NoStop}%
\bibitem [{\citenamefont {Schlamminger}\ \emph {et~al.}(2008)\citenamefont
  {Schlamminger}, \citenamefont {Choi}, \citenamefont {Wagner}, \citenamefont
  {Gundlach},\ and\ \citenamefont {Adelberger}}]{Schlamminger:2007ht}%
  \BibitemOpen
  \bibfield  {author} {\bibinfo {author} {\bibfnamefont {S.}~\bibnamefont
  {Schlamminger}}, \bibinfo {author} {\bibfnamefont {K.~Y.}\ \bibnamefont
  {Choi}}, \bibinfo {author} {\bibfnamefont {T.~A.}\ \bibnamefont {Wagner}},
  \bibinfo {author} {\bibfnamefont {J.~H.}\ \bibnamefont {Gundlach}},\ and\
  \bibinfo {author} {\bibfnamefont {E.~G.}\ \bibnamefont {Adelberger}},\
  }\bibfield  {title} {\bibinfo {title} {{Test of the equivalence principle
  using a rotating torsion balance}},\ }\href
  {https://doi.org/10.1103/PhysRevLett.100.041101} {\bibfield  {journal}
  {\bibinfo  {journal} {Phys. Rev. Lett.}\ }\textbf {\bibinfo {volume} {100}},\
  \bibinfo {pages} {041101} (\bibinfo {year} {2008})},\ \Eprint
  {https://arxiv.org/abs/0712.0607} {arXiv:0712.0607 [gr-qc]} \BibitemShut
  {NoStop}%
\bibitem [{\citenamefont {Wagner}\ \emph {et~al.}(2012)\citenamefont {Wagner},
  \citenamefont {Schlamminger}, \citenamefont {Gundlach},\ and\ \citenamefont
  {Adelberger}}]{Wagner:2012ui}%
  \BibitemOpen
  \bibfield  {author} {\bibinfo {author} {\bibfnamefont {T.~A.}\ \bibnamefont
  {Wagner}}, \bibinfo {author} {\bibfnamefont {S.}~\bibnamefont
  {Schlamminger}}, \bibinfo {author} {\bibfnamefont {J.~H.}\ \bibnamefont
  {Gundlach}},\ and\ \bibinfo {author} {\bibfnamefont {E.~G.}\ \bibnamefont
  {Adelberger}},\ }\bibfield  {title} {\bibinfo {title} {{Torsion-balance tests
  of the weak equivalence principle}},\ }\href
  {https://doi.org/10.1088/0264-9381/29/18/184002} {\bibfield  {journal}
  {\bibinfo  {journal} {Class. Quant. Grav.}\ }\textbf {\bibinfo {volume}
  {29}},\ \bibinfo {pages} {184002} (\bibinfo {year} {2012})},\ \Eprint
  {https://arxiv.org/abs/1207.2442} {arXiv:1207.2442 [gr-qc]} \BibitemShut
  {NoStop}%
\bibitem [{\citenamefont {Abbott}\ \emph {et~al.}(2021)\citenamefont {Abbott}
  \emph {et~al.}}]{Abbott:2021npy}%
  \BibitemOpen
  \bibfield  {author} {\bibinfo {author} {\bibfnamefont {R.}~\bibnamefont
  {Abbott}} \emph {et~al.} (\bibinfo {collaboration} {LIGO Scientific, Virgo,
  KAGRA}),\ }\bibfield  {title} {\bibinfo {title} {{Constraints on dark photon
  dark matter using data from LIGO's and Virgo's third observing run}},\
  }\href@noop {} {\  (\bibinfo {year} {2021})},\ \Eprint
  {https://arxiv.org/abs/2105.13085} {arXiv:2105.13085 [astro-ph.CO]}
  \BibitemShut {NoStop}%
\bibitem [{\citenamefont {Robertson}\ \emph {et~al.}(2017)\citenamefont
  {Robertson}, \citenamefont {Massey},\ and\ \citenamefont
  {Eke}}]{Robertson:2016xjh}%
  \BibitemOpen
  \bibfield  {author} {\bibinfo {author} {\bibfnamefont {A.}~\bibnamefont
  {Robertson}}, \bibinfo {author} {\bibfnamefont {R.}~\bibnamefont {Massey}},\
  and\ \bibinfo {author} {\bibfnamefont {V.}~\bibnamefont {Eke}},\ }\bibfield
  {title} {\bibinfo {title} {{What does the Bullet Cluster tell us about
  self-interacting dark matter?}},\ }\href
  {https://doi.org/10.1093/mnras/stw2670} {\bibfield  {journal} {\bibinfo
  {journal} {Mon. Not. Roy. Astron. Soc.}\ }\textbf {\bibinfo {volume} {465}},\
  \bibinfo {pages} {569} (\bibinfo {year} {2017})},\ \Eprint
  {https://arxiv.org/abs/1605.04307} {arXiv:1605.04307 [astro-ph.CO]}
  \BibitemShut {NoStop}%
\bibitem [{\citenamefont {Bodmer}(1971)}]{Bodmer:1971we}%
  \BibitemOpen
  \bibfield  {author} {\bibinfo {author} {\bibfnamefont {A.~R.}\ \bibnamefont
  {Bodmer}},\ }\bibfield  {title} {\bibinfo {title} {{Collapsed nuclei}},\
  }\href {https://doi.org/10.1103/PhysRevD.4.1601} {\bibfield  {journal}
  {\bibinfo  {journal} {Phys. Rev. D}\ }\textbf {\bibinfo {volume} {4}},\
  \bibinfo {pages} {1601} (\bibinfo {year} {1971})}\BibitemShut {NoStop}%
\bibitem [{\citenamefont {Witten}(1984)}]{Witten:1984rs}%
  \BibitemOpen
  \bibfield  {author} {\bibinfo {author} {\bibfnamefont {E.}~\bibnamefont
  {Witten}},\ }\bibfield  {title} {\bibinfo {title} {{Cosmic Separation of
  Phases}},\ }\href {https://doi.org/10.1103/PhysRevD.30.272} {\bibfield
  {journal} {\bibinfo  {journal} {Phys. Rev. D}\ }\textbf {\bibinfo {volume}
  {30}},\ \bibinfo {pages} {272} (\bibinfo {year} {1984})}\BibitemShut
  {NoStop}%
\bibitem [{\citenamefont {Bai}\ \emph {et~al.}(2019)\citenamefont {Bai},
  \citenamefont {Long},\ and\ \citenamefont {Lu}}]{Bai:2018dxf}%
  \BibitemOpen
  \bibfield  {author} {\bibinfo {author} {\bibfnamefont {Y.}~\bibnamefont
  {Bai}}, \bibinfo {author} {\bibfnamefont {A.~J.}\ \bibnamefont {Long}},\ and\
  \bibinfo {author} {\bibfnamefont {S.}~\bibnamefont {Lu}},\ }\bibfield
  {title} {\bibinfo {title} {{Dark Quark Nuggets}},\ }\href
  {https://doi.org/10.1103/PhysRevD.99.055047} {\bibfield  {journal} {\bibinfo
  {journal} {Phys. Rev. D}\ }\textbf {\bibinfo {volume} {99}},\ \bibinfo
  {pages} {055047} (\bibinfo {year} {2019})},\ \Eprint
  {https://arxiv.org/abs/1810.04360} {arXiv:1810.04360 [hep-ph]} \BibitemShut
  {NoStop}%
\bibitem [{\citenamefont {Hardy}\ \emph
  {et~al.}(2015{\natexlab{a}})\citenamefont {Hardy}, \citenamefont {Lasenby},
  \citenamefont {March-Russell},\ and\ \citenamefont {West}}]{Hardy:2014mqa}%
  \BibitemOpen
  \bibfield  {author} {\bibinfo {author} {\bibfnamefont {E.}~\bibnamefont
  {Hardy}}, \bibinfo {author} {\bibfnamefont {R.}~\bibnamefont {Lasenby}},
  \bibinfo {author} {\bibfnamefont {J.}~\bibnamefont {March-Russell}},\ and\
  \bibinfo {author} {\bibfnamefont {S.~M.}\ \bibnamefont {West}},\ }\bibfield
  {title} {\bibinfo {title} {{Big Bang Synthesis of Nuclear Dark Matter}},\
  }\href {https://doi.org/10.1007/JHEP06(2015)011} {\bibfield  {journal}
  {\bibinfo  {journal} {JHEP}\ }\textbf {\bibinfo {volume} {06}},\ \bibinfo
  {pages} {011}},\ \Eprint {https://arxiv.org/abs/1411.3739} {arXiv:1411.3739
  [hep-ph]} \BibitemShut {NoStop}%
\bibitem [{\citenamefont {Hardy}\ \emph
  {et~al.}(2015{\natexlab{b}})\citenamefont {Hardy}, \citenamefont {Lasenby},
  \citenamefont {March-Russell},\ and\ \citenamefont {West}}]{Hardy:2015boa}%
  \BibitemOpen
  \bibfield  {author} {\bibinfo {author} {\bibfnamefont {E.}~\bibnamefont
  {Hardy}}, \bibinfo {author} {\bibfnamefont {R.}~\bibnamefont {Lasenby}},
  \bibinfo {author} {\bibfnamefont {J.}~\bibnamefont {March-Russell}},\ and\
  \bibinfo {author} {\bibfnamefont {S.~M.}\ \bibnamefont {West}},\ }\bibfield
  {title} {\bibinfo {title} {{Signatures of Large Composite Dark Matter
  States}},\ }\href {https://doi.org/10.1007/JHEP07(2015)133} {\bibfield
  {journal} {\bibinfo  {journal} {JHEP}\ }\textbf {\bibinfo {volume} {07}},\
  \bibinfo {pages} {133}},\ \Eprint {https://arxiv.org/abs/1504.05419}
  {arXiv:1504.05419 [hep-ph]} \BibitemShut {NoStop}%
\bibitem [{\citenamefont {Gresham}\ \emph {et~al.}(2017)\citenamefont
  {Gresham}, \citenamefont {Lou},\ and\ \citenamefont
  {Zurek}}]{Gresham:2017zqi}%
  \BibitemOpen
  \bibfield  {author} {\bibinfo {author} {\bibfnamefont {M.~I.}\ \bibnamefont
  {Gresham}}, \bibinfo {author} {\bibfnamefont {H.~K.}\ \bibnamefont {Lou}},\
  and\ \bibinfo {author} {\bibfnamefont {K.~M.}\ \bibnamefont {Zurek}},\
  }\bibfield  {title} {\bibinfo {title} {{Nuclear Structure of Bound States of
  Asymmetric Dark Matter}},\ }\href
  {https://doi.org/10.1103/PhysRevD.96.096012} {\bibfield  {journal} {\bibinfo
  {journal} {Phys. Rev. D}\ }\textbf {\bibinfo {volume} {96}},\ \bibinfo
  {pages} {096012} (\bibinfo {year} {2017})},\ \Eprint
  {https://arxiv.org/abs/1707.02313} {arXiv:1707.02313 [hep-ph]} \BibitemShut
  {NoStop}%
\bibitem [{\citenamefont {Redi}\ and\ \citenamefont
  {Tesi}(2019)}]{Redi:2018muu}%
  \BibitemOpen
  \bibfield  {author} {\bibinfo {author} {\bibfnamefont {M.}~\bibnamefont
  {Redi}}\ and\ \bibinfo {author} {\bibfnamefont {A.}~\bibnamefont {Tesi}},\
  }\bibfield  {title} {\bibinfo {title} {{Cosmological Production of Dark
  Nuclei}},\ }\href {https://doi.org/10.1007/JHEP04(2019)108} {\bibfield
  {journal} {\bibinfo  {journal} {JHEP}\ }\textbf {\bibinfo {volume} {04}},\
  \bibinfo {pages} {108}},\ \Eprint {https://arxiv.org/abs/1812.08784}
  {arXiv:1812.08784 [hep-ph]} \BibitemShut {NoStop}%
\bibitem [{\citenamefont {Detmold}\ \emph
  {et~al.}(2014{\natexlab{a}})\citenamefont {Detmold}, \citenamefont
  {McCullough},\ and\ \citenamefont {Pochinsky}}]{Detmold:2014qqa}%
  \BibitemOpen
  \bibfield  {author} {\bibinfo {author} {\bibfnamefont {W.}~\bibnamefont
  {Detmold}}, \bibinfo {author} {\bibfnamefont {M.}~\bibnamefont
  {McCullough}},\ and\ \bibinfo {author} {\bibfnamefont {A.}~\bibnamefont
  {Pochinsky}},\ }\bibfield  {title} {\bibinfo {title} {{Dark Nuclei I:
  Cosmology and Indirect Detection}},\ }\href
  {https://doi.org/10.1103/PhysRevD.90.115013} {\bibfield  {journal} {\bibinfo
  {journal} {Phys. Rev. D}\ }\textbf {\bibinfo {volume} {90}},\ \bibinfo
  {pages} {115013} (\bibinfo {year} {2014}{\natexlab{a}})},\ \Eprint
  {https://arxiv.org/abs/1406.2276} {arXiv:1406.2276 [hep-ph]} \BibitemShut
  {NoStop}%
\bibitem [{\citenamefont {Detmold}\ \emph
  {et~al.}(2014{\natexlab{b}})\citenamefont {Detmold}, \citenamefont
  {McCullough},\ and\ \citenamefont {Pochinsky}}]{Detmold:2014kba}%
  \BibitemOpen
  \bibfield  {author} {\bibinfo {author} {\bibfnamefont {W.}~\bibnamefont
  {Detmold}}, \bibinfo {author} {\bibfnamefont {M.}~\bibnamefont
  {McCullough}},\ and\ \bibinfo {author} {\bibfnamefont {A.}~\bibnamefont
  {Pochinsky}},\ }\bibfield  {title} {\bibinfo {title} {{Dark nuclei. II.
  Nuclear spectroscopy in two-color QCD}},\ }\href
  {https://doi.org/10.1103/PhysRevD.90.114506} {\bibfield  {journal} {\bibinfo
  {journal} {Phys. Rev. D}\ }\textbf {\bibinfo {volume} {90}},\ \bibinfo
  {pages} {114506} (\bibinfo {year} {2014}{\natexlab{b}})},\ \Eprint
  {https://arxiv.org/abs/1406.4116} {arXiv:1406.4116 [hep-lat]} \BibitemShut
  {NoStop}%
\bibitem [{\citenamefont {Gresham}\ \emph
  {et~al.}(2018{\natexlab{a}})\citenamefont {Gresham}, \citenamefont {Lou},\
  and\ \citenamefont {Zurek}}]{Gresham:2018anj}%
  \BibitemOpen
  \bibfield  {author} {\bibinfo {author} {\bibfnamefont {M.~I.}\ \bibnamefont
  {Gresham}}, \bibinfo {author} {\bibfnamefont {H.~K.}\ \bibnamefont {Lou}},\
  and\ \bibinfo {author} {\bibfnamefont {K.~M.}\ \bibnamefont {Zurek}},\
  }\bibfield  {title} {\bibinfo {title} {{Astrophysical Signatures of
  Asymmetric Dark Matter Bound States}},\ }\href
  {https://doi.org/10.1103/PhysRevD.98.096001} {\bibfield  {journal} {\bibinfo
  {journal} {Phys. Rev. D}\ }\textbf {\bibinfo {volume} {98}},\ \bibinfo
  {pages} {096001} (\bibinfo {year} {2018}{\natexlab{a}})},\ \Eprint
  {https://arxiv.org/abs/1805.04512} {arXiv:1805.04512 [hep-ph]} \BibitemShut
  {NoStop}%
\bibitem [{\citenamefont {Krnjaic}\ and\ \citenamefont
  {Sigurdson}(2015)}]{Krnjaic:2014xza}%
  \BibitemOpen
  \bibfield  {author} {\bibinfo {author} {\bibfnamefont {G.}~\bibnamefont
  {Krnjaic}}\ and\ \bibinfo {author} {\bibfnamefont {K.}~\bibnamefont
  {Sigurdson}},\ }\bibfield  {title} {\bibinfo {title} {{Big Bang
  Darkleosynthesis}},\ }\href {https://doi.org/10.1016/j.physletb.2015.11.001}
  {\bibfield  {journal} {\bibinfo  {journal} {Phys. Lett. B}\ }\textbf
  {\bibinfo {volume} {751}},\ \bibinfo {pages} {464} (\bibinfo {year}
  {2015})},\ \Eprint {https://arxiv.org/abs/1406.1171} {arXiv:1406.1171
  [hep-ph]} \BibitemShut {NoStop}%
\bibitem [{\citenamefont {Gresham}\ \emph
  {et~al.}(2018{\natexlab{b}})\citenamefont {Gresham}, \citenamefont {Lou},\
  and\ \citenamefont {Zurek}}]{Gresham:2017cvl}%
  \BibitemOpen
  \bibfield  {author} {\bibinfo {author} {\bibfnamefont {M.~I.}\ \bibnamefont
  {Gresham}}, \bibinfo {author} {\bibfnamefont {H.~K.}\ \bibnamefont {Lou}},\
  and\ \bibinfo {author} {\bibfnamefont {K.~M.}\ \bibnamefont {Zurek}},\
  }\bibfield  {title} {\bibinfo {title} {{Early Universe synthesis of
  asymmetric dark matter nuggets}},\ }\href
  {https://doi.org/10.1103/PhysRevD.97.036003} {\bibfield  {journal} {\bibinfo
  {journal} {Phys. Rev. D}\ }\textbf {\bibinfo {volume} {97}},\ \bibinfo
  {pages} {036003} (\bibinfo {year} {2018}{\natexlab{b}})},\ \Eprint
  {https://arxiv.org/abs/1707.02316} {arXiv:1707.02316 [hep-ph]} \BibitemShut
  {NoStop}%
\bibitem [{\citenamefont {Enqvist}\ \emph {et~al.}(2002)\citenamefont
  {Enqvist}, \citenamefont {Jokinen}, \citenamefont {Multamaki},\ and\
  \citenamefont {Vilja}}]{Enqvist:2001jd}%
  \BibitemOpen
  \bibfield  {author} {\bibinfo {author} {\bibfnamefont {K.}~\bibnamefont
  {Enqvist}}, \bibinfo {author} {\bibfnamefont {A.}~\bibnamefont {Jokinen}},
  \bibinfo {author} {\bibfnamefont {T.}~\bibnamefont {Multamaki}},\ and\
  \bibinfo {author} {\bibfnamefont {I.}~\bibnamefont {Vilja}},\ }\bibfield
  {title} {\bibinfo {title} {{Constraints on selfinteracting Q ball dark
  matter}},\ }\href {https://doi.org/10.1016/S0370-2693(01)01500-3} {\bibfield
  {journal} {\bibinfo  {journal} {Phys. Lett. B}\ }\textbf {\bibinfo {volume}
  {526}},\ \bibinfo {pages} {9} (\bibinfo {year} {2002})},\ \Eprint
  {https://arxiv.org/abs/hep-ph/0111348} {arXiv:hep-ph/0111348} \BibitemShut
  {NoStop}%
\bibitem [{\citenamefont {Cotner}\ and\ \citenamefont
  {Kusenko}(2016)}]{cotner_kusenko_2016}%
  \BibitemOpen
  \bibfield  {author} {\bibinfo {author} {\bibfnamefont {E.}~\bibnamefont
  {Cotner}}\ and\ \bibinfo {author} {\bibfnamefont {A.}~\bibnamefont
  {Kusenko}},\ }\bibfield  {title} {\bibinfo {title} {Astrophysical constraints
  on dark-matter q-balls in the presence of baryon-violating operators},\
  }\bibfield  {journal} {\bibinfo  {journal} {Physical Review D}\ }\textbf
  {\bibinfo {volume} {94}},\ \href {https://doi.org/10.1103/physrevd.94.123006}
  {10.1103/physrevd.94.123006} (\bibinfo {year} {2016})\BibitemShut {NoStop}%
\bibitem [{\citenamefont {Chung}\ \emph
  {et~al.}(1998{\natexlab{a}})\citenamefont {Chung}, \citenamefont {Kolb},\
  and\ \citenamefont {Riotto}}]{Chung:1998zb}%
  \BibitemOpen
  \bibfield  {author} {\bibinfo {author} {\bibfnamefont {D.~J.~H.}\
  \bibnamefont {Chung}}, \bibinfo {author} {\bibfnamefont {E.~W.}\ \bibnamefont
  {Kolb}},\ and\ \bibinfo {author} {\bibfnamefont {A.}~\bibnamefont {Riotto}},\
  }\bibfield  {title} {\bibinfo {title} {{Superheavy dark matter}},\ }\href
  {https://doi.org/10.1103/PhysRevD.59.023501} {\bibfield  {journal} {\bibinfo
  {journal} {Phys. Rev. D}\ }\textbf {\bibinfo {volume} {59}},\ \bibinfo
  {pages} {023501} (\bibinfo {year} {1998}{\natexlab{a}})},\ \Eprint
  {https://arxiv.org/abs/hep-ph/9802238} {arXiv:hep-ph/9802238} \BibitemShut
  {NoStop}%
\bibitem [{\citenamefont {Chung}\ \emph
  {et~al.}(1998{\natexlab{b}})\citenamefont {Chung}, \citenamefont {Kolb},\
  and\ \citenamefont {Riotto}}]{Chung:1998ua}%
  \BibitemOpen
  \bibfield  {author} {\bibinfo {author} {\bibfnamefont {D.~J.~H.}\
  \bibnamefont {Chung}}, \bibinfo {author} {\bibfnamefont {E.~W.}\ \bibnamefont
  {Kolb}},\ and\ \bibinfo {author} {\bibfnamefont {A.}~\bibnamefont {Riotto}},\
  }\bibfield  {title} {\bibinfo {title} {{Nonthermal supermassive dark
  matter}},\ }\href {https://doi.org/10.1103/PhysRevLett.81.4048} {\bibfield
  {journal} {\bibinfo  {journal} {Phys. Rev. Lett.}\ }\textbf {\bibinfo
  {volume} {81}},\ \bibinfo {pages} {4048} (\bibinfo {year}
  {1998}{\natexlab{b}})},\ \Eprint {https://arxiv.org/abs/hep-ph/9805473}
  {arXiv:hep-ph/9805473} \BibitemShut {NoStop}%
\bibitem [{\citenamefont {Bernal}\ and\ \citenamefont
  {Zapata}(2021{\natexlab{a}})}]{Bernal:2020bjf}%
  \BibitemOpen
  \bibfield  {author} {\bibinfo {author} {\bibfnamefont {N.}~\bibnamefont
  {Bernal}}\ and\ \bibinfo {author} {\bibfnamefont {O.}~\bibnamefont
  {Zapata}},\ }\bibfield  {title} {\bibinfo {title} {{Dark Matter in the Time
  of Primordial Black Holes}},\ }\href
  {https://doi.org/10.1088/1475-7516/2021/03/015} {\bibfield  {journal}
  {\bibinfo  {journal} {JCAP}\ }\textbf {\bibinfo {volume} {03}},\ \bibinfo
  {pages} {015}},\ \Eprint {https://arxiv.org/abs/2011.12306} {arXiv:2011.12306
  [astro-ph.CO]} \BibitemShut {NoStop}%
\bibitem [{\citenamefont {Bai}\ and\ \citenamefont
  {Orlofsky}(2020)}]{Bai:2019zcd}%
  \BibitemOpen
  \bibfield  {author} {\bibinfo {author} {\bibfnamefont {Y.}~\bibnamefont
  {Bai}}\ and\ \bibinfo {author} {\bibfnamefont {N.}~\bibnamefont {Orlofsky}},\
  }\bibfield  {title} {\bibinfo {title} {{Primordial Extremal Black Holes as
  Dark Matter}},\ }\href {https://doi.org/10.1103/PhysRevD.101.055006}
  {\bibfield  {journal} {\bibinfo  {journal} {Phys. Rev. D}\ }\textbf {\bibinfo
  {volume} {101}},\ \bibinfo {pages} {055006} (\bibinfo {year} {2020})},\
  \Eprint {https://arxiv.org/abs/1906.04858} {arXiv:1906.04858 [hep-ph]}
  \BibitemShut {NoStop}%
\bibitem [{\citenamefont {Aydemir}\ \emph {et~al.}(2020)\citenamefont
  {Aydemir}, \citenamefont {Holdom},\ and\ \citenamefont
  {Ren}}]{Aydemir:2020xfd}%
  \BibitemOpen
  \bibfield  {author} {\bibinfo {author} {\bibfnamefont {U.}~\bibnamefont
  {Aydemir}}, \bibinfo {author} {\bibfnamefont {B.}~\bibnamefont {Holdom}},\
  and\ \bibinfo {author} {\bibfnamefont {J.}~\bibnamefont {Ren}},\ }\bibfield
  {title} {\bibinfo {title} {{Not quite black holes as dark matter}},\ }\href
  {https://doi.org/10.1103/PhysRevD.102.024058} {\bibfield  {journal} {\bibinfo
   {journal} {Phys. Rev. D}\ }\textbf {\bibinfo {volume} {102}},\ \bibinfo
  {pages} {024058} (\bibinfo {year} {2020})},\ \Eprint
  {https://arxiv.org/abs/2003.10682} {arXiv:2003.10682 [gr-qc]} \BibitemShut
  {NoStop}%
\bibitem [{\citenamefont {MacGibbon}(1987)}]{MacGibbon:1987my}%
  \BibitemOpen
  \bibfield  {author} {\bibinfo {author} {\bibfnamefont {J.~H.}\ \bibnamefont
  {MacGibbon}},\ }\bibfield  {title} {\bibinfo {title} {{Can Planck-mass relics
  of evaporating black holes close the universe?}},\ }\href
  {https://doi.org/10.1038/329308a0} {\bibfield  {journal} {\bibinfo  {journal}
  {Nature}\ }\textbf {\bibinfo {volume} {329}},\ \bibinfo {pages} {308}
  (\bibinfo {year} {1987})}\BibitemShut {NoStop}%
\bibitem [{\citenamefont {Salvio}\ and\ \citenamefont
  {Veerm\"ae}(2020)}]{Salvio:2019llz}%
  \BibitemOpen
  \bibfield  {author} {\bibinfo {author} {\bibfnamefont {A.}~\bibnamefont
  {Salvio}}\ and\ \bibinfo {author} {\bibfnamefont {H.}~\bibnamefont
  {Veerm\"ae}},\ }\bibfield  {title} {\bibinfo {title} {{Horizonless
  ultracompact objects and dark matter in quadratic gravity}},\ }\href
  {https://doi.org/10.1088/1475-7516/2020/02/018} {\bibfield  {journal}
  {\bibinfo  {journal} {JCAP}\ }\textbf {\bibinfo {volume} {02}},\ \bibinfo
  {pages} {018}},\ \Eprint {https://arxiv.org/abs/1912.13333} {arXiv:1912.13333
  [gr-qc]} \BibitemShut {NoStop}%
\bibitem [{\citenamefont {Evans}\ \emph {et~al.}(2019)\citenamefont {Evans},
  \citenamefont {O'Hare},\ and\ \citenamefont {McCabe}}]{Evans:2018bqy}%
  \BibitemOpen
  \bibfield  {author} {\bibinfo {author} {\bibfnamefont {N.~W.}\ \bibnamefont
  {Evans}}, \bibinfo {author} {\bibfnamefont {C.~A.~J.}\ \bibnamefont
  {O'Hare}},\ and\ \bibinfo {author} {\bibfnamefont {C.}~\bibnamefont
  {McCabe}},\ }\bibfield  {title} {\bibinfo {title} {{Refinement of the
  standard halo model for dark matter searches in light of the Gaia Sausage}},\
  }\href {https://doi.org/10.1103/PhysRevD.99.023012} {\bibfield  {journal}
  {\bibinfo  {journal} {Phys. Rev. D}\ }\textbf {\bibinfo {volume} {99}},\
  \bibinfo {pages} {023012} (\bibinfo {year} {2019})},\ \Eprint
  {https://arxiv.org/abs/1810.11468} {arXiv:1810.11468 [astro-ph.GA]}
  \BibitemShut {NoStop}%
\bibitem [{\citenamefont {Coleman}(1985)}]{Coleman:1985ki}%
  \BibitemOpen
  \bibfield  {author} {\bibinfo {author} {\bibfnamefont {S.~R.}\ \bibnamefont
  {Coleman}},\ }\bibfield  {title} {\bibinfo {title} {{Q Balls}},\ }\href
  {https://doi.org/10.1016/0550-3213(86)90520-1} {\bibfield  {journal}
  {\bibinfo  {journal} {Nucl. Phys. B}\ }\textbf {\bibinfo {volume} {262}},\
  \bibinfo {pages} {263} (\bibinfo {year} {1985})},\ \bibinfo {note} {[Erratum:
  Nucl.Phys.B 269, 744 (1986)]}\BibitemShut {NoStop}%
\bibitem [{\citenamefont {Friedberg}\ \emph {et~al.}(1976)\citenamefont
  {Friedberg}, \citenamefont {Lee},\ and\ \citenamefont
  {Sirlin}}]{Friedberg:1976me}%
  \BibitemOpen
  \bibfield  {author} {\bibinfo {author} {\bibfnamefont {R.}~\bibnamefont
  {Friedberg}}, \bibinfo {author} {\bibfnamefont {T.~D.}\ \bibnamefont {Lee}},\
  and\ \bibinfo {author} {\bibfnamefont {A.}~\bibnamefont {Sirlin}},\
  }\bibfield  {title} {\bibinfo {title} {{A Class of Scalar-Field Soliton
  Solutions in Three Space Dimensions}},\ }\href
  {https://doi.org/10.1103/PhysRevD.13.2739} {\bibfield  {journal} {\bibinfo
  {journal} {Phys. Rev. D}\ }\textbf {\bibinfo {volume} {13}},\ \bibinfo
  {pages} {2739} (\bibinfo {year} {1976})}\BibitemShut {NoStop}%
\bibitem [{\citenamefont {Kusenko}\ and\ \citenamefont
  {Shaposhnikov}(1998)}]{Kusenko:1997si}%
  \BibitemOpen
  \bibfield  {author} {\bibinfo {author} {\bibfnamefont {A.}~\bibnamefont
  {Kusenko}}\ and\ \bibinfo {author} {\bibfnamefont {M.~E.}\ \bibnamefont
  {Shaposhnikov}},\ }\bibfield  {title} {\bibinfo {title} {{Supersymmetric Q
  balls as dark matter}},\ }\href
  {https://doi.org/10.1016/S0370-2693(97)01375-0} {\bibfield  {journal}
  {\bibinfo  {journal} {Phys. Lett. B}\ }\textbf {\bibinfo {volume} {418}},\
  \bibinfo {pages} {46} (\bibinfo {year} {1998})},\ \Eprint
  {https://arxiv.org/abs/hep-ph/9709492} {arXiv:hep-ph/9709492} \BibitemShut
  {NoStop}%
\bibitem [{\citenamefont {Kusenko}(1997)}]{Kusenko:1997zq}%
  \BibitemOpen
  \bibfield  {author} {\bibinfo {author} {\bibfnamefont {A.}~\bibnamefont
  {Kusenko}},\ }\bibfield  {title} {\bibinfo {title} {{Solitons in the
  supersymmetric extensions of the standard model}},\ }\href
  {https://doi.org/10.1016/S0370-2693(97)00584-4} {\bibfield  {journal}
  {\bibinfo  {journal} {Phys. Lett. B}\ }\textbf {\bibinfo {volume} {405}},\
  \bibinfo {pages} {108} (\bibinfo {year} {1997})},\ \Eprint
  {https://arxiv.org/abs/hep-ph/9704273} {arXiv:hep-ph/9704273} \BibitemShut
  {NoStop}%
\bibitem [{\citenamefont {Kusenko}(2006)}]{Kusenko:2006gv}%
  \BibitemOpen
  \bibfield  {author} {\bibinfo {author} {\bibfnamefont {A.}~\bibnamefont
  {Kusenko}},\ }\bibfield  {title} {\bibinfo {title} {{Properties and
  signatures of supersymmetric Q-balls}},\ }in\ \href@noop {} {\emph {\bibinfo
  {booktitle} {{Workshop on Exotic Physics with Neutrino Telescopes}}}}\
  (\bibinfo {year} {2006})\ \Eprint {https://arxiv.org/abs/hep-ph/0612159}
  {arXiv:hep-ph/0612159} \BibitemShut {NoStop}%
\bibitem [{\citenamefont {Hong}\ \emph {et~al.}(2016)\citenamefont {Hong},
  \citenamefont {Kawasaki},\ and\ \citenamefont {Yamada}}]{Hong:2016ict}%
  \BibitemOpen
  \bibfield  {author} {\bibinfo {author} {\bibfnamefont {J.-P.}\ \bibnamefont
  {Hong}}, \bibinfo {author} {\bibfnamefont {M.}~\bibnamefont {Kawasaki}},\
  and\ \bibinfo {author} {\bibfnamefont {M.}~\bibnamefont {Yamada}},\
  }\bibfield  {title} {\bibinfo {title} {{Charged Q-ball Dark Matter from $B$
  and $L$ direction}},\ }\href {https://doi.org/10.1088/1475-7516/2016/08/053}
  {\bibfield  {journal} {\bibinfo  {journal} {JCAP}\ }\textbf {\bibinfo
  {volume} {08}},\ \bibinfo {pages} {053}},\ \Eprint
  {https://arxiv.org/abs/1604.04352} {arXiv:1604.04352 [hep-ph]} \BibitemShut
  {NoStop}%
\bibitem [{\citenamefont {Hong}\ \emph {et~al.}(2020)\citenamefont {Hong},
  \citenamefont {Jung},\ and\ \citenamefont {Xie}}]{Hong:2020est}%
  \BibitemOpen
  \bibfield  {author} {\bibinfo {author} {\bibfnamefont {J.-P.}\ \bibnamefont
  {Hong}}, \bibinfo {author} {\bibfnamefont {S.}~\bibnamefont {Jung}},\ and\
  \bibinfo {author} {\bibfnamefont {K.-P.}\ \bibnamefont {Xie}},\ }\bibfield
  {title} {\bibinfo {title} {{Fermi-ball dark matter from a first-order phase
  transition}},\ }\href {https://doi.org/10.1103/PhysRevD.102.075028}
  {\bibfield  {journal} {\bibinfo  {journal} {Phys. Rev. D}\ }\textbf {\bibinfo
  {volume} {102}},\ \bibinfo {pages} {075028} (\bibinfo {year} {2020})},\
  \Eprint {https://arxiv.org/abs/2008.04430} {arXiv:2008.04430 [hep-ph]}
  \BibitemShut {NoStop}%
\bibitem [{\citenamefont {Holdom}(1987{\natexlab{a}})}]{Holdom:1987ep}%
  \BibitemOpen
  \bibfield  {author} {\bibinfo {author} {\bibfnamefont {B.}~\bibnamefont
  {Holdom}},\ }\bibfield  {title} {\bibinfo {title} {{Cosmic Balls of Trapped
  Neutrinos}},\ }\href {https://doi.org/10.1103/PhysRevD.36.1000} {\bibfield
  {journal} {\bibinfo  {journal} {Phys. Rev. D}\ }\textbf {\bibinfo {volume}
  {36}},\ \bibinfo {pages} {1000} (\bibinfo {year}
  {1987}{\natexlab{a}})}\BibitemShut {NoStop}%
\bibitem [{\citenamefont {Lohiya}(1994)}]{Lohiya:1994pmf}%
  \BibitemOpen
  \bibfield  {author} {\bibinfo {author} {\bibfnamefont {D.}~\bibnamefont
  {Lohiya}},\ }\bibfield  {title} {\bibinfo {title} {{Nontopological solitons
  in nonminimally coupled scalar fields: Theory and consequences}},\
  }\href@noop {} {\  (\bibinfo {year} {1994})},\ \Eprint
  {https://arxiv.org/abs/gr-qc/9407014} {arXiv:gr-qc/9407014} \BibitemShut
  {NoStop}%
\bibitem [{\citenamefont {Stojkovic}(2003)}]{Stojkovic:2001qi}%
  \BibitemOpen
  \bibfield  {author} {\bibinfo {author} {\bibfnamefont {D.}~\bibnamefont
  {Stojkovic}},\ }\bibfield  {title} {\bibinfo {title} {{Nontopological
  solitons in brane world models}},\ }\href
  {https://doi.org/10.1103/PhysRevD.67.045012} {\bibfield  {journal} {\bibinfo
  {journal} {Phys. Rev. D}\ }\textbf {\bibinfo {volume} {67}},\ \bibinfo
  {pages} {045012} (\bibinfo {year} {2003})},\ \Eprint
  {https://arxiv.org/abs/hep-ph/0111061} {arXiv:hep-ph/0111061} \BibitemShut
  {NoStop}%
\bibitem [{\citenamefont {Macpherson}\ and\ \citenamefont
  {Campbell}(1995)}]{Macpherson:1994wf}%
  \BibitemOpen
  \bibfield  {author} {\bibinfo {author} {\bibfnamefont {A.~L.}\ \bibnamefont
  {Macpherson}}\ and\ \bibinfo {author} {\bibfnamefont {B.~A.}\ \bibnamefont
  {Campbell}},\ }\bibfield  {title} {\bibinfo {title} {{Biased discrete
  symmetry breaking and Fermi balls}},\ }\href
  {https://doi.org/10.1016/0370-2693(95)00080-5} {\bibfield  {journal}
  {\bibinfo  {journal} {Phys. Lett. B}\ }\textbf {\bibinfo {volume} {347}},\
  \bibinfo {pages} {205} (\bibinfo {year} {1995})},\ \Eprint
  {https://arxiv.org/abs/hep-ph/9408387} {arXiv:hep-ph/9408387} \BibitemShut
  {NoStop}%
\bibitem [{\citenamefont {Holdom}(1987{\natexlab{b}})}]{Holdom:1987bu}%
  \BibitemOpen
  \bibfield  {author} {\bibinfo {author} {\bibfnamefont {B.}~\bibnamefont
  {Holdom}},\ }\bibfield  {title} {\bibinfo {title} {{COSMIC NEUTRINO BALLS}},\
  }\href@noop {} {\  (\bibinfo {year} {1987}{\natexlab{b}})}\BibitemShut
  {NoStop}%
\bibitem [{\citenamefont {Zhitnitsky}(2003)}]{Zhitnitsky:2002qa}%
  \BibitemOpen
  \bibfield  {author} {\bibinfo {author} {\bibfnamefont {A.~R.}\ \bibnamefont
  {Zhitnitsky}},\ }\bibfield  {title} {\bibinfo {title} {{'Nonbaryonic' dark
  matter as baryonic color superconductor}},\ }\href
  {https://doi.org/10.1088/1475-7516/2003/10/010} {\bibfield  {journal}
  {\bibinfo  {journal} {JCAP}\ }\textbf {\bibinfo {volume} {10}},\ \bibinfo
  {pages} {010}},\ \Eprint {https://arxiv.org/abs/hep-ph/0202161}
  {arXiv:hep-ph/0202161} \BibitemShut {NoStop}%
\bibitem [{\citenamefont {Ogure}\ \emph {et~al.}(2003)\citenamefont {Ogure},
  \citenamefont {Yoshida},\ and\ \citenamefont {Arafune}}]{Ogure:2002hv}%
  \BibitemOpen
  \bibfield  {author} {\bibinfo {author} {\bibfnamefont {K.}~\bibnamefont
  {Ogure}}, \bibinfo {author} {\bibfnamefont {T.}~\bibnamefont {Yoshida}},\
  and\ \bibinfo {author} {\bibfnamefont {J.}~\bibnamefont {Arafune}},\
  }\bibfield  {title} {\bibinfo {title} {{Stable neutral Fermi ball}},\ }\href
  {https://doi.org/10.1103/PhysRevD.67.123518} {\bibfield  {journal} {\bibinfo
  {journal} {Phys. Rev. D}\ }\textbf {\bibinfo {volume} {67}},\ \bibinfo
  {pages} {123518} (\bibinfo {year} {2003})},\ \Eprint
  {https://arxiv.org/abs/hep-ph/0212332} {arXiv:hep-ph/0212332} \BibitemShut
  {NoStop}%
\bibitem [{\citenamefont {Frieman}\ \emph {et~al.}(1988)\citenamefont
  {Frieman}, \citenamefont {Gelmini}, \citenamefont {Gleiser},\ and\
  \citenamefont {Kolb}}]{Frieman:1988ut}%
  \BibitemOpen
  \bibfield  {author} {\bibinfo {author} {\bibfnamefont {J.~A.}\ \bibnamefont
  {Frieman}}, \bibinfo {author} {\bibfnamefont {G.~B.}\ \bibnamefont
  {Gelmini}}, \bibinfo {author} {\bibfnamefont {M.}~\bibnamefont {Gleiser}},\
  and\ \bibinfo {author} {\bibfnamefont {E.~W.}\ \bibnamefont {Kolb}},\
  }\bibfield  {title} {\bibinfo {title} {{Solitogenesis: Primordial Origin of
  Nontopological Solitons}},\ }\href
  {https://doi.org/10.1103/PhysRevLett.60.2101} {\bibfield  {journal} {\bibinfo
   {journal} {Phys. Rev. Lett.}\ }\textbf {\bibinfo {volume} {60}},\ \bibinfo
  {pages} {2101} (\bibinfo {year} {1988})}\BibitemShut {NoStop}%
\bibitem [{\citenamefont {Affleck}\ and\ \citenamefont
  {Dine}(1985)}]{Affleck:1984fy}%
  \BibitemOpen
  \bibfield  {author} {\bibinfo {author} {\bibfnamefont {I.}~\bibnamefont
  {Affleck}}\ and\ \bibinfo {author} {\bibfnamefont {M.}~\bibnamefont {Dine}},\
  }\bibfield  {title} {\bibinfo {title} {{A New Mechanism for Baryogenesis}},\
  }\href {https://doi.org/10.1016/0550-3213(85)90021-5} {\bibfield  {journal}
  {\bibinfo  {journal} {Nucl. Phys. B}\ }\textbf {\bibinfo {volume} {249}},\
  \bibinfo {pages} {361} (\bibinfo {year} {1985})}\BibitemShut {NoStop}%
\bibitem [{\citenamefont {Dine}\ and\ \citenamefont
  {Kusenko}(2003)}]{Dine:2003ax}%
  \BibitemOpen
  \bibfield  {author} {\bibinfo {author} {\bibfnamefont {M.}~\bibnamefont
  {Dine}}\ and\ \bibinfo {author} {\bibfnamefont {A.}~\bibnamefont {Kusenko}},\
  }\bibfield  {title} {\bibinfo {title} {{The Origin of the matter - antimatter
  asymmetry}},\ }\href {https://doi.org/10.1103/RevModPhys.76.1} {\bibfield
  {journal} {\bibinfo  {journal} {Rev. Mod. Phys.}\ }\textbf {\bibinfo {volume}
  {76}},\ \bibinfo {pages} {1} (\bibinfo {year} {2003})},\ \Eprint
  {https://arxiv.org/abs/hep-ph/0303065} {arXiv:hep-ph/0303065} \BibitemShut
  {NoStop}%
\bibitem [{\citenamefont {Krylov}\ \emph {et~al.}(2013)\citenamefont {Krylov},
  \citenamefont {Levin},\ and\ \citenamefont {Rubakov}}]{Krylov:2013qe}%
  \BibitemOpen
  \bibfield  {author} {\bibinfo {author} {\bibfnamefont {E.}~\bibnamefont
  {Krylov}}, \bibinfo {author} {\bibfnamefont {A.}~\bibnamefont {Levin}},\ and\
  \bibinfo {author} {\bibfnamefont {V.}~\bibnamefont {Rubakov}},\ }\bibfield
  {title} {\bibinfo {title} {{Cosmological phase transition, baryon asymmetry
  and dark matter Q-balls}},\ }\href
  {https://doi.org/10.1103/PhysRevD.87.083528} {\bibfield  {journal} {\bibinfo
  {journal} {Phys. Rev. D}\ }\textbf {\bibinfo {volume} {87}},\ \bibinfo
  {pages} {083528} (\bibinfo {year} {2013})},\ \Eprint
  {https://arxiv.org/abs/1301.0354} {arXiv:1301.0354 [hep-ph]} \BibitemShut
  {NoStop}%
\bibitem [{\citenamefont {Kolb}\ \emph {et~al.}(1999)\citenamefont {Kolb},
  \citenamefont {Chung},\ and\ \citenamefont {Riotto}}]{Kolb:1998ki}%
  \BibitemOpen
  \bibfield  {author} {\bibinfo {author} {\bibfnamefont {E.~W.}\ \bibnamefont
  {Kolb}}, \bibinfo {author} {\bibfnamefont {D.~J.~H.}\ \bibnamefont {Chung}},\
  and\ \bibinfo {author} {\bibfnamefont {A.}~\bibnamefont {Riotto}},\
  }\bibfield  {title} {\bibinfo {title} {{WIMPzillas!}},\ }\href
  {https://doi.org/10.1063/1.59655} {\bibfield  {journal} {\bibinfo  {journal}
  {AIP Conf. Proc.}\ }\textbf {\bibinfo {volume} {484}},\ \bibinfo {pages} {91}
  (\bibinfo {year} {1999})},\ \Eprint {https://arxiv.org/abs/hep-ph/9810361}
  {arXiv:hep-ph/9810361} \BibitemShut {NoStop}%
\bibitem [{\citenamefont {Garny}\ \emph {et~al.}(2016)\citenamefont {Garny},
  \citenamefont {Sandora},\ and\ \citenamefont {Sloth}}]{Garny:2015sjg}%
  \BibitemOpen
  \bibfield  {author} {\bibinfo {author} {\bibfnamefont {M.}~\bibnamefont
  {Garny}}, \bibinfo {author} {\bibfnamefont {M.}~\bibnamefont {Sandora}},\
  and\ \bibinfo {author} {\bibfnamefont {M.~S.}\ \bibnamefont {Sloth}},\
  }\bibfield  {title} {\bibinfo {title} {{Planckian Interacting Massive
  Particles as Dark Matter}},\ }\href
  {https://doi.org/10.1103/PhysRevLett.116.101302} {\bibfield  {journal}
  {\bibinfo  {journal} {Phys. Rev. Lett.}\ }\textbf {\bibinfo {volume} {116}},\
  \bibinfo {pages} {101302} (\bibinfo {year} {2016})},\ \Eprint
  {https://arxiv.org/abs/1511.03278} {arXiv:1511.03278 [hep-ph]} \BibitemShut
  {NoStop}%
\bibitem [{\citenamefont {Kolb}\ and\ \citenamefont
  {Long}(2017)}]{Kolb:2017jvz}%
  \BibitemOpen
  \bibfield  {author} {\bibinfo {author} {\bibfnamefont {E.~W.}\ \bibnamefont
  {Kolb}}\ and\ \bibinfo {author} {\bibfnamefont {A.~J.}\ \bibnamefont
  {Long}},\ }\bibfield  {title} {\bibinfo {title} {{Superheavy dark matter
  through Higgs portal operators}},\ }\href
  {https://doi.org/10.1103/PhysRevD.96.103540} {\bibfield  {journal} {\bibinfo
  {journal} {Phys. Rev. D}\ }\textbf {\bibinfo {volume} {96}},\ \bibinfo
  {pages} {103540} (\bibinfo {year} {2017})},\ \Eprint
  {https://arxiv.org/abs/1708.04293} {arXiv:1708.04293 [astro-ph.CO]}
  \BibitemShut {NoStop}%
\bibitem [{\citenamefont {Harigaya}\ \emph {et~al.}(2016)\citenamefont
  {Harigaya}, \citenamefont {Lin},\ and\ \citenamefont
  {Lou}}]{Harigaya:2016vda}%
  \BibitemOpen
  \bibfield  {author} {\bibinfo {author} {\bibfnamefont {K.}~\bibnamefont
  {Harigaya}}, \bibinfo {author} {\bibfnamefont {T.}~\bibnamefont {Lin}},\ and\
  \bibinfo {author} {\bibfnamefont {H.~K.}\ \bibnamefont {Lou}},\ }\bibfield
  {title} {\bibinfo {title} {{GUTzilla Dark Matter}},\ }\href
  {https://doi.org/10.1007/JHEP09(2016)014} {\bibfield  {journal} {\bibinfo
  {journal} {JHEP}\ }\textbf {\bibinfo {volume} {09}},\ \bibinfo {pages}
  {014}},\ \Eprint {https://arxiv.org/abs/1606.00923} {arXiv:1606.00923
  [hep-ph]} \BibitemShut {NoStop}%
\bibitem [{\citenamefont {Berlin}(2017)}]{Berlin:2017ife}%
  \BibitemOpen
  \bibfield  {author} {\bibinfo {author} {\bibfnamefont {A.}~\bibnamefont
  {Berlin}},\ }\bibfield  {title} {\bibinfo {title} {{WIMPs with GUTs: Dark
  Matter Coannihilation with a Lighter Species}},\ }\href
  {https://doi.org/10.1103/PhysRevLett.119.121801} {\bibfield  {journal}
  {\bibinfo  {journal} {Phys. Rev. Lett.}\ }\textbf {\bibinfo {volume} {119}},\
  \bibinfo {pages} {121801} (\bibinfo {year} {2017})},\ \Eprint
  {https://arxiv.org/abs/1704.08256} {arXiv:1704.08256 [hep-ph]} \BibitemShut
  {NoStop}%
\bibitem [{\citenamefont {Chung}\ \emph {et~al.}(2000)\citenamefont {Chung},
  \citenamefont {Kolb}, \citenamefont {Riotto},\ and\ \citenamefont
  {Tkachev}}]{Chung:1999ve}%
  \BibitemOpen
  \bibfield  {author} {\bibinfo {author} {\bibfnamefont {D.~J.~H.}\
  \bibnamefont {Chung}}, \bibinfo {author} {\bibfnamefont {E.~W.}\ \bibnamefont
  {Kolb}}, \bibinfo {author} {\bibfnamefont {A.}~\bibnamefont {Riotto}},\ and\
  \bibinfo {author} {\bibfnamefont {I.~I.}\ \bibnamefont {Tkachev}},\
  }\bibfield  {title} {\bibinfo {title} {{Probing Planckian physics: Resonant
  production of particles during inflation and features in the primordial power
  spectrum}},\ }\href {https://doi.org/10.1103/PhysRevD.62.043508} {\bibfield
  {journal} {\bibinfo  {journal} {Phys. Rev. D}\ }\textbf {\bibinfo {volume}
  {62}},\ \bibinfo {pages} {043508} (\bibinfo {year} {2000})},\ \Eprint
  {https://arxiv.org/abs/hep-ph/9910437} {arXiv:hep-ph/9910437} \BibitemShut
  {NoStop}%
\bibitem [{\citenamefont {Fedderke}\ \emph {et~al.}(2015)\citenamefont
  {Fedderke}, \citenamefont {Kolb},\ and\ \citenamefont
  {Wyman}}]{Fedderke:2014ura}%
  \BibitemOpen
  \bibfield  {author} {\bibinfo {author} {\bibfnamefont {M.~A.}\ \bibnamefont
  {Fedderke}}, \bibinfo {author} {\bibfnamefont {E.~W.}\ \bibnamefont {Kolb}},\
  and\ \bibinfo {author} {\bibfnamefont {M.}~\bibnamefont {Wyman}},\ }\bibfield
   {title} {\bibinfo {title} {{Irruption of massive particle species during
  inflation}},\ }\href {https://doi.org/10.1103/PhysRevD.91.063505} {\bibfield
  {journal} {\bibinfo  {journal} {Phys. Rev. D}\ }\textbf {\bibinfo {volume}
  {91}},\ \bibinfo {pages} {063505} (\bibinfo {year} {2015})},\ \Eprint
  {https://arxiv.org/abs/1409.1584} {arXiv:1409.1584 [astro-ph.CO]}
  \BibitemShut {NoStop}%
\bibitem [{\citenamefont {Kuzmin}\ and\ \citenamefont
  {Tkachev}(1999)}]{Kuzmin:1998kk}%
  \BibitemOpen
  \bibfield  {author} {\bibinfo {author} {\bibfnamefont {V.}~\bibnamefont
  {Kuzmin}}\ and\ \bibinfo {author} {\bibfnamefont {I.}~\bibnamefont
  {Tkachev}},\ }\bibfield  {title} {\bibinfo {title} {{Matter creation via
  vacuum fluctuations in the early universe and observed ultrahigh-energy
  cosmic ray events}},\ }\href {https://doi.org/10.1103/PhysRevD.59.123006}
  {\bibfield  {journal} {\bibinfo  {journal} {Phys. Rev. D}\ }\textbf {\bibinfo
  {volume} {59}},\ \bibinfo {pages} {123006} (\bibinfo {year} {1999})},\
  \Eprint {https://arxiv.org/abs/hep-ph/9809547} {arXiv:hep-ph/9809547}
  \BibitemShut {NoStop}%
\bibitem [{\citenamefont {Park}\ and\ \citenamefont
  {Park}(2014)}]{park_park_2014}%
  \BibitemOpen
  \bibfield  {author} {\bibinfo {author} {\bibfnamefont {J.-C.}\ \bibnamefont
  {Park}}\ and\ \bibinfo {author} {\bibfnamefont {S.~C.}\ \bibnamefont
  {Park}},\ }\bibfield  {title} {\bibinfo {title} {A testable scenario of
  wimpzilla with dark radiation},\ }\href
  {https://doi.org/10.1016/j.physletb.2013.11.027} {\bibfield  {journal}
  {\bibinfo  {journal} {Physics Letters B}\ }\textbf {\bibinfo {volume}
  {728}},\ \bibinfo {pages} {41–44} (\bibinfo {year} {2014})}\BibitemShut
  {NoStop}%
\bibitem [{\citenamefont {Chung}\ \emph {et~al.}(2001)\citenamefont {Chung},
  \citenamefont {Crotty}, \citenamefont {Kolb},\ and\ \citenamefont
  {Riotto}}]{chung_crotty_kolb_riotto_2001}%
  \BibitemOpen
  \bibfield  {author} {\bibinfo {author} {\bibfnamefont {D.~J.~H.}\
  \bibnamefont {Chung}}, \bibinfo {author} {\bibfnamefont {P.}~\bibnamefont
  {Crotty}}, \bibinfo {author} {\bibfnamefont {E.~W.}\ \bibnamefont {Kolb}},\
  and\ \bibinfo {author} {\bibfnamefont {A.}~\bibnamefont {Riotto}},\
  }\bibfield  {title} {\bibinfo {title} {Gravitational production of superheavy
  dark matter},\ }\bibfield  {journal} {\bibinfo  {journal} {Physical Review
  D}\ }\textbf {\bibinfo {volume} {64}},\ \href
  {https://doi.org/10.1103/physrevd.64.043503} {10.1103/physrevd.64.043503}
  (\bibinfo {year} {2001})\BibitemShut {NoStop}%
\bibitem [{\citenamefont {Farzinnia}\ and\ \citenamefont
  {Kouwn}(2016)}]{farzinnia_kouwn_2016}%
  \BibitemOpen
  \bibfield  {author} {\bibinfo {author} {\bibfnamefont {A.}~\bibnamefont
  {Farzinnia}}\ and\ \bibinfo {author} {\bibfnamefont {S.}~\bibnamefont
  {Kouwn}},\ }\bibfield  {title} {\bibinfo {title} {Classically scale invariant
  inflation, supermassive wimps, and adimensional gravity},\ }\bibfield
  {journal} {\bibinfo  {journal} {Physical Review D}\ }\textbf {\bibinfo
  {volume} {93}},\ \href {https://doi.org/10.1103/physrevd.93.063528}
  {10.1103/physrevd.93.063528} (\bibinfo {year} {2016})\BibitemShut {NoStop}%
\bibitem [{\citenamefont {Li}\ \emph {et~al.}(2020)\citenamefont {Li},
  \citenamefont {Lu}, \citenamefont {Wang},\ and\ \citenamefont
  {Zhou}}]{li_lu_wang_zhou_2020}%
  \BibitemOpen
  \bibfield  {author} {\bibinfo {author} {\bibfnamefont {L.}~\bibnamefont
  {Li}}, \bibinfo {author} {\bibfnamefont {S.}~\bibnamefont {Lu}}, \bibinfo
  {author} {\bibfnamefont {Y.}~\bibnamefont {Wang}},\ and\ \bibinfo {author}
  {\bibfnamefont {S.}~\bibnamefont {Zhou}},\ }\bibfield  {title} {\bibinfo
  {title} {Cosmological signatures of superheavy dark matter},\ }\bibfield
  {journal} {\bibinfo  {journal} {Journal of High Energy Physics}\ }\textbf
  {\bibinfo {volume} {2020}},\ \href {https://doi.org/10.1007/jhep07(2020)231}
  {10.1007/jhep07(2020)231} (\bibinfo {year} {2020})\BibitemShut {NoStop}%
\bibitem [{\citenamefont {Bernal}\ and\ \citenamefont
  {Zapata}(2021{\natexlab{b}})}]{bernal_zapata_2021}%
  \BibitemOpen
  \bibfield  {author} {\bibinfo {author} {\bibfnamefont {N.}~\bibnamefont
  {Bernal}}\ and\ \bibinfo {author} {\bibfnamefont {O.}~\bibnamefont
  {Zapata}},\ }\bibfield  {title} {\bibinfo {title} {Self-interacting dark
  matter from primordial black holes},\ }\href
  {https://doi.org/10.1088/1475-7516/2021/03/007} {\bibfield  {journal}
  {\bibinfo  {journal} {Journal of Cosmology and Astroparticle Physics}\
  }\textbf {\bibinfo {volume} {2021}}\bibinfo  {number} { (03)},\ \bibinfo
  {pages} {007}}\BibitemShut {NoStop}%
\bibitem [{\citenamefont {Gondolo}\ \emph {et~al.}(2020)\citenamefont
  {Gondolo}, \citenamefont {Sandick},\ and\ \citenamefont
  {Haghi}}]{gondolo_sandick_haghi_2020}%
  \BibitemOpen
\bibfield  {number} {  }\bibfield  {author} {\bibinfo {author} {\bibfnamefont
  {P.}~\bibnamefont {Gondolo}}, \bibinfo {author} {\bibfnamefont
  {P.}~\bibnamefont {Sandick}},\ and\ \bibinfo {author} {\bibfnamefont
  {B.~S.~E.}\ \bibnamefont {Haghi}},\ }\bibfield  {title} {\bibinfo {title}
  {Effects of primordial black holes on dark matter models},\ }\bibfield
  {journal} {\bibinfo  {journal} {Physical Review D}\ }\textbf {\bibinfo
  {volume} {102}},\ \href {https://doi.org/10.1103/physrevd.102.095018}
  {10.1103/physrevd.102.095018} (\bibinfo {year} {2020})\BibitemShut {NoStop}%
\bibitem [{\citenamefont {De~Rujula}\ and\ \citenamefont
  {Glashow}(1984)}]{DeRujula:1984axn}%
  \BibitemOpen
  \bibfield  {author} {\bibinfo {author} {\bibfnamefont {A.}~\bibnamefont
  {De~Rujula}}\ and\ \bibinfo {author} {\bibfnamefont {S.~L.}\ \bibnamefont
  {Glashow}},\ }\bibfield  {title} {\bibinfo {title} {{Nuclearites: A Novel
  Form of Cosmic Radiation}},\ }\href {https://doi.org/10.1038/312734a0}
  {\bibfield  {journal} {\bibinfo  {journal} {Nature}\ }\textbf {\bibinfo
  {volume} {312}},\ \bibinfo {pages} {734} (\bibinfo {year}
  {1984})}\BibitemShut {NoStop}%
\bibitem [{\citenamefont {Kusenko}\ and\ \citenamefont
  {Steinhardt}(2001)}]{Kusenko:2001vu}%
  \BibitemOpen
  \bibfield  {author} {\bibinfo {author} {\bibfnamefont {A.}~\bibnamefont
  {Kusenko}}\ and\ \bibinfo {author} {\bibfnamefont {P.~J.}\ \bibnamefont
  {Steinhardt}},\ }\bibfield  {title} {\bibinfo {title} {{Q ball candidates for
  selfinteracting dark matter}},\ }\href
  {https://doi.org/10.1103/PhysRevLett.87.141301} {\bibfield  {journal}
  {\bibinfo  {journal} {Phys. Rev. Lett.}\ }\textbf {\bibinfo {volume} {87}},\
  \bibinfo {pages} {141301} (\bibinfo {year} {2001})},\ \Eprint
  {https://arxiv.org/abs/astro-ph/0106008} {arXiv:astro-ph/0106008}
  \BibitemShut {NoStop}%
\bibitem [{\citenamefont {Spergel}\ and\ \citenamefont
  {Steinhardt}(2000)}]{Spergel:1999mh}%
  \BibitemOpen
  \bibfield  {author} {\bibinfo {author} {\bibfnamefont {D.~N.}\ \bibnamefont
  {Spergel}}\ and\ \bibinfo {author} {\bibfnamefont {P.~J.}\ \bibnamefont
  {Steinhardt}},\ }\bibfield  {title} {\bibinfo {title} {{Observational
  evidence for selfinteracting cold dark matter}},\ }\href
  {https://doi.org/10.1103/PhysRevLett.84.3760} {\bibfield  {journal} {\bibinfo
   {journal} {Phys. Rev. Lett.}\ }\textbf {\bibinfo {volume} {84}},\ \bibinfo
  {pages} {3760} (\bibinfo {year} {2000})},\ \Eprint
  {https://arxiv.org/abs/astro-ph/9909386} {arXiv:astro-ph/9909386}
  \BibitemShut {NoStop}%
\bibitem [{\citenamefont {Jackson~Kimball}\ \emph {et~al.}(2018)\citenamefont
  {Jackson~Kimball}, \citenamefont {Budker}, \citenamefont {Eby}, \citenamefont
  {Pospelov}, \citenamefont {Pustelny}, \citenamefont {Scholtes}, \citenamefont
  {Stadnik}, \citenamefont {Weis},\ and\ \citenamefont
  {Wickenbrock}}]{JacksonKimball:2017qgk}%
  \BibitemOpen
  \bibfield  {author} {\bibinfo {author} {\bibfnamefont {D.~F.}\ \bibnamefont
  {Jackson~Kimball}}, \bibinfo {author} {\bibfnamefont {D.}~\bibnamefont
  {Budker}}, \bibinfo {author} {\bibfnamefont {J.}~\bibnamefont {Eby}},
  \bibinfo {author} {\bibfnamefont {M.}~\bibnamefont {Pospelov}}, \bibinfo
  {author} {\bibfnamefont {S.}~\bibnamefont {Pustelny}}, \bibinfo {author}
  {\bibfnamefont {T.}~\bibnamefont {Scholtes}}, \bibinfo {author}
  {\bibfnamefont {Y.~V.}\ \bibnamefont {Stadnik}}, \bibinfo {author}
  {\bibfnamefont {A.}~\bibnamefont {Weis}},\ and\ \bibinfo {author}
  {\bibfnamefont {A.}~\bibnamefont {Wickenbrock}},\ }\bibfield  {title}
  {\bibinfo {title} {{Searching for axion stars and Q-balls with a terrestrial
  magnetometer network}},\ }\href {https://doi.org/10.1103/PhysRevD.97.043002}
  {\bibfield  {journal} {\bibinfo  {journal} {Phys. Rev. D}\ }\textbf {\bibinfo
  {volume} {97}},\ \bibinfo {pages} {043002} (\bibinfo {year} {2018})},\
  \Eprint {https://arxiv.org/abs/1710.04323} {arXiv:1710.04323
  [physics.atom-ph]} \BibitemShut {NoStop}%
\bibitem [{\citenamefont {Jacobs}\ \emph {et~al.}(2015)\citenamefont {Jacobs},
  \citenamefont {Starkman},\ and\ \citenamefont {Lynn}}]{Jacobs:2014yca}%
  \BibitemOpen
  \bibfield  {author} {\bibinfo {author} {\bibfnamefont {D.~M.}\ \bibnamefont
  {Jacobs}}, \bibinfo {author} {\bibfnamefont {G.~D.}\ \bibnamefont
  {Starkman}},\ and\ \bibinfo {author} {\bibfnamefont {B.~W.}\ \bibnamefont
  {Lynn}},\ }\bibfield  {title} {\bibinfo {title} {{Macro Dark Matter}},\
  }\href {https://doi.org/10.1093/mnras/stv774} {\bibfield  {journal} {\bibinfo
   {journal} {Mon. Not. Roy. Astron. Soc.}\ }\textbf {\bibinfo {volume}
  {450}},\ \bibinfo {pages} {3418} (\bibinfo {year} {2015})},\ \Eprint
  {https://arxiv.org/abs/1410.2236} {arXiv:1410.2236 [astro-ph.CO]}
  \BibitemShut {NoStop}%
\bibitem [{\citenamefont {Singh~Sidhu}\ \emph {et~al.}(2020)\citenamefont
  {Singh~Sidhu}, \citenamefont {Scherrer},\ and\ \citenamefont
  {Starkman}}]{SinghSidhu:2019loh}%
  \BibitemOpen
  \bibfield  {author} {\bibinfo {author} {\bibfnamefont {J.}~\bibnamefont
  {Singh~Sidhu}}, \bibinfo {author} {\bibfnamefont {R.~J.}\ \bibnamefont
  {Scherrer}},\ and\ \bibinfo {author} {\bibfnamefont {G.}~\bibnamefont
  {Starkman}},\ }\bibfield  {title} {\bibinfo {title} {{Death and serious
  injury from dark matter}},\ }\href
  {https://doi.org/10.1016/j.physletb.2020.135300} {\bibfield  {journal}
  {\bibinfo  {journal} {Phys. Lett. B}\ }\textbf {\bibinfo {volume} {803}},\
  \bibinfo {pages} {135300} (\bibinfo {year} {2020})},\ \Eprint
  {https://arxiv.org/abs/1907.06674} {arXiv:1907.06674 [astro-ph.CO]}
  \BibitemShut {NoStop}%
\bibitem [{\citenamefont {Singh~Sidhu}(2020)}]{SinghSidhu:2019nmh}%
  \BibitemOpen
  \bibfield  {author} {\bibinfo {author} {\bibfnamefont {J.}~\bibnamefont
  {Singh~Sidhu}},\ }\bibfield  {title} {\bibinfo {title} {{Charge Constraints
  of Macroscopic Dark Matter}},\ }\href
  {https://doi.org/10.1103/PhysRevD.101.043526} {\bibfield  {journal} {\bibinfo
   {journal} {Phys. Rev. D}\ }\textbf {\bibinfo {volume} {101}},\ \bibinfo
  {pages} {043526} (\bibinfo {year} {2020})},\ \Eprint
  {https://arxiv.org/abs/1912.04732} {arXiv:1912.04732 [astro-ph.CO]}
  \BibitemShut {NoStop}%
\bibitem [{\citenamefont {Clark}\ \emph {et~al.}(2020)\citenamefont {Clark},
  \citenamefont {Depoian}, \citenamefont {Elshimy}, \citenamefont {Kopec},
  \citenamefont {Lang},\ and\ \citenamefont {Qin}}]{Clark:2020mna}%
  \BibitemOpen
  \bibfield  {author} {\bibinfo {author} {\bibfnamefont {M.}~\bibnamefont
  {Clark}}, \bibinfo {author} {\bibfnamefont {A.}~\bibnamefont {Depoian}},
  \bibinfo {author} {\bibfnamefont {B.}~\bibnamefont {Elshimy}}, \bibinfo
  {author} {\bibfnamefont {A.}~\bibnamefont {Kopec}}, \bibinfo {author}
  {\bibfnamefont {R.~F.}\ \bibnamefont {Lang}},\ and\ \bibinfo {author}
  {\bibfnamefont {J.}~\bibnamefont {Qin}},\ }\bibfield  {title} {\bibinfo
  {title} {{Direct Detection Limits on Heavy Dark Matter}},\ }\href
  {https://doi.org/10.1103/PhysRevD.102.123026} {\bibfield  {journal} {\bibinfo
   {journal} {Phys. Rev. D}\ }\textbf {\bibinfo {volume} {102}},\ \bibinfo
  {pages} {123026} (\bibinfo {year} {2020})},\ \Eprint
  {https://arxiv.org/abs/2009.07909} {arXiv:2009.07909 [hep-ph]} \BibitemShut
  {NoStop}%
\bibitem [{\citenamefont {Digman}\ \emph {et~al.}(2019)\citenamefont {Digman},
  \citenamefont {Cappiello}, \citenamefont {Beacom}, \citenamefont {Hirata},\
  and\ \citenamefont {Peter}}]{Digman:2019wdm}%
  \BibitemOpen
  \bibfield  {author} {\bibinfo {author} {\bibfnamefont {M.~C.}\ \bibnamefont
  {Digman}}, \bibinfo {author} {\bibfnamefont {C.~V.}\ \bibnamefont
  {Cappiello}}, \bibinfo {author} {\bibfnamefont {J.~F.}\ \bibnamefont
  {Beacom}}, \bibinfo {author} {\bibfnamefont {C.~M.}\ \bibnamefont {Hirata}},\
  and\ \bibinfo {author} {\bibfnamefont {A.~H.~G.}\ \bibnamefont {Peter}},\
  }\bibfield  {title} {\bibinfo {title} {{Not as big as a barn: Upper bounds on
  dark matter-nucleus cross sections}},\ }\href
  {https://doi.org/10.1103/PhysRevD.100.063013} {\bibfield  {journal} {\bibinfo
   {journal} {Phys. Rev. D}\ }\textbf {\bibinfo {volume} {100}},\ \bibinfo
  {pages} {063013} (\bibinfo {year} {2019})},\ \Eprint
  {https://arxiv.org/abs/1907.10618} {arXiv:1907.10618 [hep-ph]} \BibitemShut
  {NoStop}%
\bibitem [{\citenamefont {Bhoonah}\ \emph {et~al.}(2020)\citenamefont
  {Bhoonah}, \citenamefont {Bramante}, \citenamefont {Schon},\ and\
  \citenamefont {Song}}]{Bhoonah:2020dzs}%
  \BibitemOpen
  \bibfield  {author} {\bibinfo {author} {\bibfnamefont {A.}~\bibnamefont
  {Bhoonah}}, \bibinfo {author} {\bibfnamefont {J.}~\bibnamefont {Bramante}},
  \bibinfo {author} {\bibfnamefont {S.}~\bibnamefont {Schon}},\ and\ \bibinfo
  {author} {\bibfnamefont {N.}~\bibnamefont {Song}},\ }\bibfield  {title}
  {\bibinfo {title} {{Detecting Composite Dark Matter with Long Range and
  Contact Interactions in Gas Clouds}},\ }\href@noop {} {\  (\bibinfo {year}
  {2020})},\ \Eprint {https://arxiv.org/abs/2010.07240} {arXiv:2010.07240
  [hep-ph]} \BibitemShut {NoStop}%
\bibitem [{\citenamefont {Graham}\ \emph {et~al.}(2018)\citenamefont {Graham},
  \citenamefont {Janish}, \citenamefont {Narayan}, \citenamefont {Rajendran},\
  and\ \citenamefont {Riggins}}]{Graham:2018efk}%
  \BibitemOpen
  \bibfield  {author} {\bibinfo {author} {\bibfnamefont {P.~W.}\ \bibnamefont
  {Graham}}, \bibinfo {author} {\bibfnamefont {R.}~\bibnamefont {Janish}},
  \bibinfo {author} {\bibfnamefont {V.}~\bibnamefont {Narayan}}, \bibinfo
  {author} {\bibfnamefont {S.}~\bibnamefont {Rajendran}},\ and\ \bibinfo
  {author} {\bibfnamefont {P.}~\bibnamefont {Riggins}},\ }\bibfield  {title}
  {\bibinfo {title} {{White Dwarfs as Dark Matter Detectors}},\ }\href
  {https://doi.org/10.1103/PhysRevD.98.115027} {\bibfield  {journal} {\bibinfo
  {journal} {Phys. Rev. D}\ }\textbf {\bibinfo {volume} {98}},\ \bibinfo
  {pages} {115027} (\bibinfo {year} {2018})},\ \Eprint
  {https://arxiv.org/abs/1805.07381} {arXiv:1805.07381 [hep-ph]} \BibitemShut
  {NoStop}%
\bibitem [{\citenamefont {Pont\'on}\ \emph {et~al.}(2019)\citenamefont
  {Pont\'on}, \citenamefont {Bai},\ and\ \citenamefont
  {Jain}}]{Ponton:2019hux}%
  \BibitemOpen
  \bibfield  {author} {\bibinfo {author} {\bibfnamefont {E.}~\bibnamefont
  {Pont\'on}}, \bibinfo {author} {\bibfnamefont {Y.}~\bibnamefont {Bai}},\ and\
  \bibinfo {author} {\bibfnamefont {B.}~\bibnamefont {Jain}},\ }\bibfield
  {title} {\bibinfo {title} {{Electroweak Symmetric Dark Matter Balls}},\
  }\href {https://doi.org/10.1007/s13130-019-11194-5} {\bibfield  {journal}
  {\bibinfo  {journal} {JHEP}\ }\textbf {\bibinfo {volume} {09}},\ \bibinfo
  {pages} {011}},\ \Eprint {https://arxiv.org/abs/1906.10739} {arXiv:1906.10739
  [hep-ph]} \BibitemShut {NoStop}%
\bibitem [{\citenamefont {Arbey}\ \emph {et~al.}(2020)\citenamefont {Arbey},
  \citenamefont {Auffinger},\ and\ \citenamefont {Silk}}]{Arbey:2019vqx}%
  \BibitemOpen
  \bibfield  {author} {\bibinfo {author} {\bibfnamefont {A.}~\bibnamefont
  {Arbey}}, \bibinfo {author} {\bibfnamefont {J.}~\bibnamefont {Auffinger}},\
  and\ \bibinfo {author} {\bibfnamefont {J.}~\bibnamefont {Silk}},\ }\bibfield
  {title} {\bibinfo {title} {{Constraining primordial black hole masses with
  the isotropic gamma ray background}},\ }\href
  {https://doi.org/10.1103/PhysRevD.101.023010} {\bibfield  {journal} {\bibinfo
   {journal} {Phys. Rev. D}\ }\textbf {\bibinfo {volume} {101}},\ \bibinfo
  {pages} {023010} (\bibinfo {year} {2020})},\ \Eprint
  {https://arxiv.org/abs/1906.04750} {arXiv:1906.04750 [astro-ph.CO]}
  \BibitemShut {NoStop}%
\bibitem [{\citenamefont {Villanueva-Domingo}\ \emph
  {et~al.}(2021)\citenamefont {Villanueva-Domingo}, \citenamefont {Mena},\ and\
  \citenamefont {Palomares-Ruiz}}]{Villanueva-Domingo:2021spv}%
  \BibitemOpen
  \bibfield  {author} {\bibinfo {author} {\bibfnamefont {P.}~\bibnamefont
  {Villanueva-Domingo}}, \bibinfo {author} {\bibfnamefont {O.}~\bibnamefont
  {Mena}},\ and\ \bibinfo {author} {\bibfnamefont {S.}~\bibnamefont
  {Palomares-Ruiz}},\ }\bibfield  {title} {\bibinfo {title} {{A brief review on
  primordial black holes as dark matter}},\ }\href
  {https://doi.org/10.3389/fspas.2021.681084} {\bibfield  {journal} {\bibinfo
  {journal} {Front. Astron. Space Sci.}\ }\textbf {\bibinfo {volume} {8}},\
  \bibinfo {pages} {87} (\bibinfo {year} {2021})},\ \Eprint
  {https://arxiv.org/abs/2103.12087} {arXiv:2103.12087 [astro-ph.CO]}
  \BibitemShut {NoStop}%
\bibitem [{\citenamefont {Read}(2014)}]{Read:2014qva}%
  \BibitemOpen
  \bibfield  {author} {\bibinfo {author} {\bibfnamefont {J.~I.}\ \bibnamefont
  {Read}},\ }\bibfield  {title} {\bibinfo {title} {{The Local Dark Matter
  Density}},\ }\href {https://doi.org/10.1088/0954-3899/41/6/063101} {\bibfield
   {journal} {\bibinfo  {journal} {J. Phys. G}\ }\textbf {\bibinfo {volume}
  {41}},\ \bibinfo {pages} {063101} (\bibinfo {year} {2014})},\ \Eprint
  {https://arxiv.org/abs/1404.1938} {arXiv:1404.1938 [astro-ph.GA]}
  \BibitemShut {NoStop}%
\bibitem [{\citenamefont {Pitjev}\ and\ \citenamefont
  {Pitjeva}(2013)}]{Pitjev:2013sfa}%
  \BibitemOpen
  \bibfield  {author} {\bibinfo {author} {\bibfnamefont {N.~P.}\ \bibnamefont
  {Pitjev}}\ and\ \bibinfo {author} {\bibfnamefont {E.~V.}\ \bibnamefont
  {Pitjeva}},\ }\bibfield  {title} {\bibinfo {title} {{Constraints on dark
  matter in the solar system}},\ }\href
  {https://doi.org/10.1134/S1063773713020060} {\bibfield  {journal} {\bibinfo
  {journal} {Astron. Lett.}\ }\textbf {\bibinfo {volume} {39}},\ \bibinfo
  {pages} {141} (\bibinfo {year} {2013})},\ \Eprint
  {https://arxiv.org/abs/1306.5534} {arXiv:1306.5534 [astro-ph.EP]}
  \BibitemShut {NoStop}%
\bibitem [{\citenamefont {Hooper}\ \emph {et~al.}(2019)\citenamefont {Hooper},
  \citenamefont {Krnjaic},\ and\ \citenamefont {McDermott}}]{Hooper:2019gtx}%
  \BibitemOpen
  \bibfield  {author} {\bibinfo {author} {\bibfnamefont {D.}~\bibnamefont
  {Hooper}}, \bibinfo {author} {\bibfnamefont {G.}~\bibnamefont {Krnjaic}},\
  and\ \bibinfo {author} {\bibfnamefont {S.~D.}\ \bibnamefont {McDermott}},\
  }\bibfield  {title} {\bibinfo {title} {{Dark Radiation and Superheavy Dark
  Matter from Black Hole Domination}},\ }\href
  {https://doi.org/10.1007/JHEP08(2019)001} {\bibfield  {journal} {\bibinfo
  {journal} {JHEP}\ }\textbf {\bibinfo {volume} {08}},\ \bibinfo {pages}
  {001}},\ \Eprint {https://arxiv.org/abs/1905.01301} {arXiv:1905.01301
  [hep-ph]} \BibitemShut {NoStop}%
\bibitem [{\citenamefont {Carr}\ and\ \citenamefont
  {Kuhnel}(2021)}]{Carr:2021bzv}%
  \BibitemOpen
  \bibfield  {author} {\bibinfo {author} {\bibfnamefont {B.}~\bibnamefont
  {Carr}}\ and\ \bibinfo {author} {\bibfnamefont {F.}~\bibnamefont {Kuhnel}},\
  }\bibfield  {title} {\bibinfo {title} {{Primordial Black Holes as Dark Matter
  Candidates}},\ }in\ \href@noop {} {\emph {\bibinfo {booktitle} {{Les Houches
  summer school on Dark Matter}}}}\ (\bibinfo {year} {2021})\ \Eprint
  {https://arxiv.org/abs/2110.02821} {arXiv:2110.02821 [astro-ph.CO]}
  \BibitemShut {NoStop}%
\end{thebibliography}%

\clearpage
\newpage
\section{Appendix}

\begin{figure}[H]
    \includegraphics[width=1\linewidth]{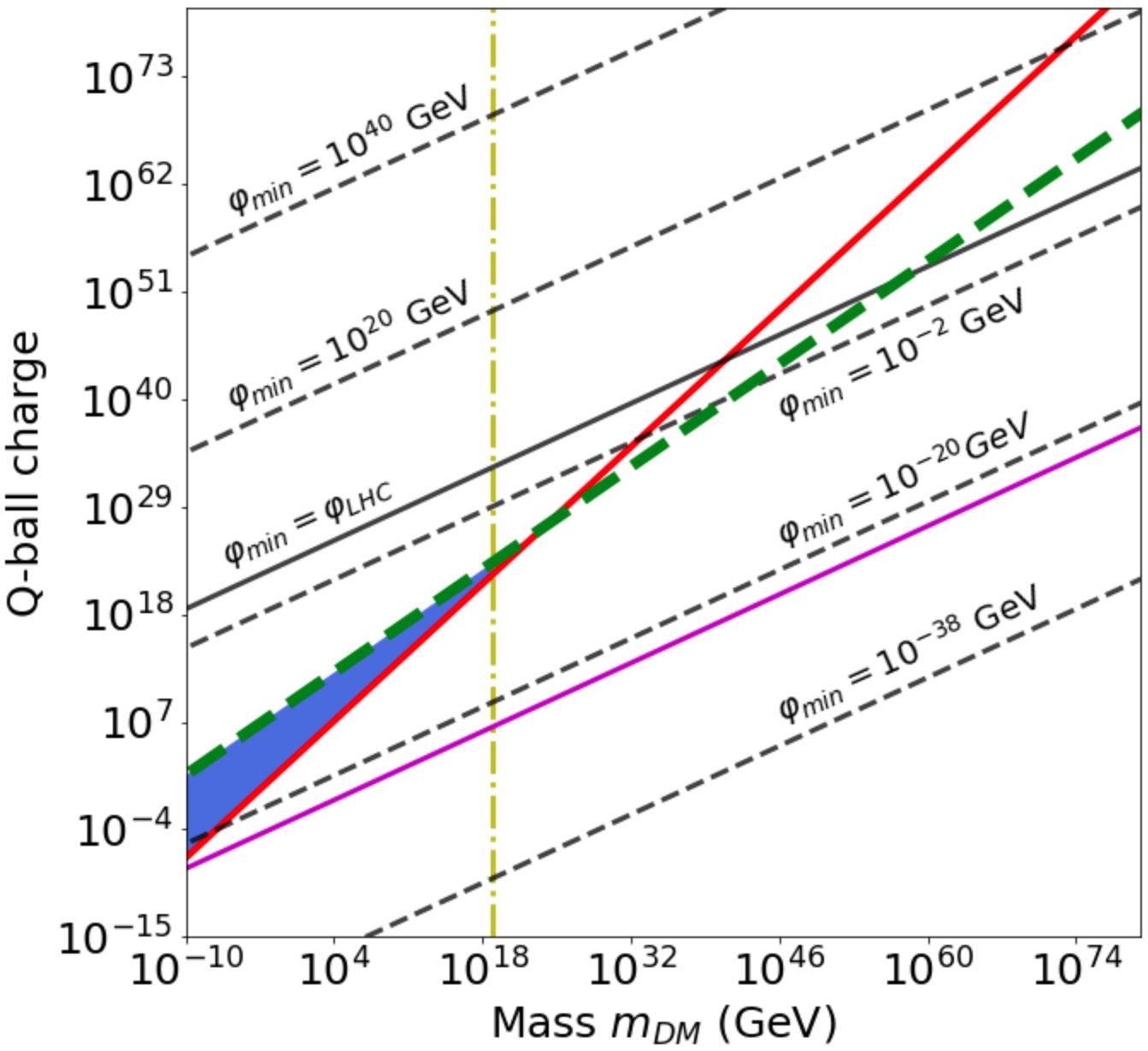}
    \caption{This is the same plot as Fig.~\ref{fig:Type1}, with the extended parameter space to show the regions excluded by the black hole limit.}
    \label{fig:plot 1}
\end{figure}
\begin{figure}[H]
    \includegraphics[width=1\linewidth]{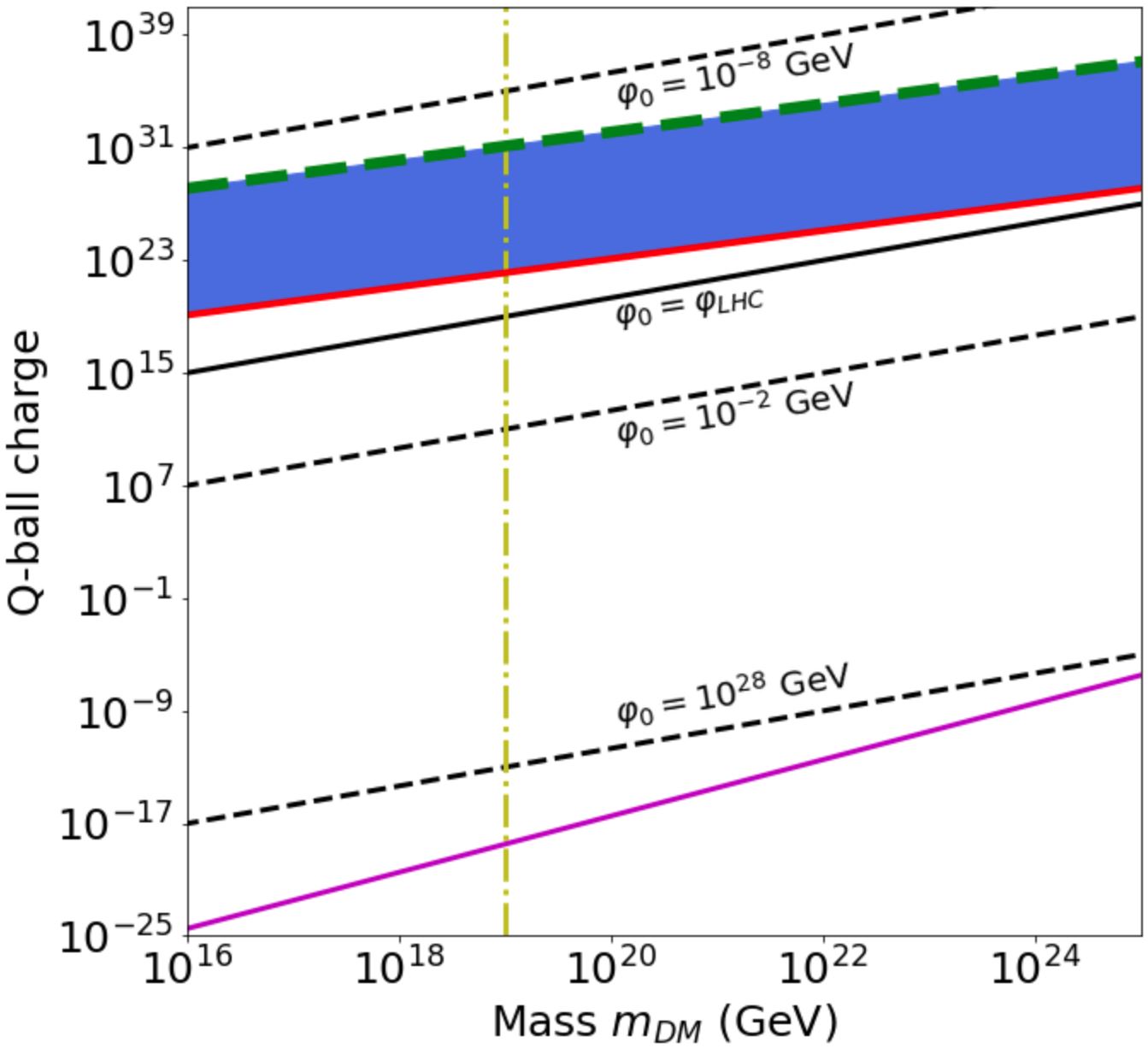}
    \caption{This is the same plot as Fig.~\ref{fig:Type2}, with the added magenta line representing the black hole limit as a minimum charge as a function of mass.}
    \label{fig:plot 2}
\end{figure}
\begin{figure}[H]
    \includegraphics[width=1\linewidth]{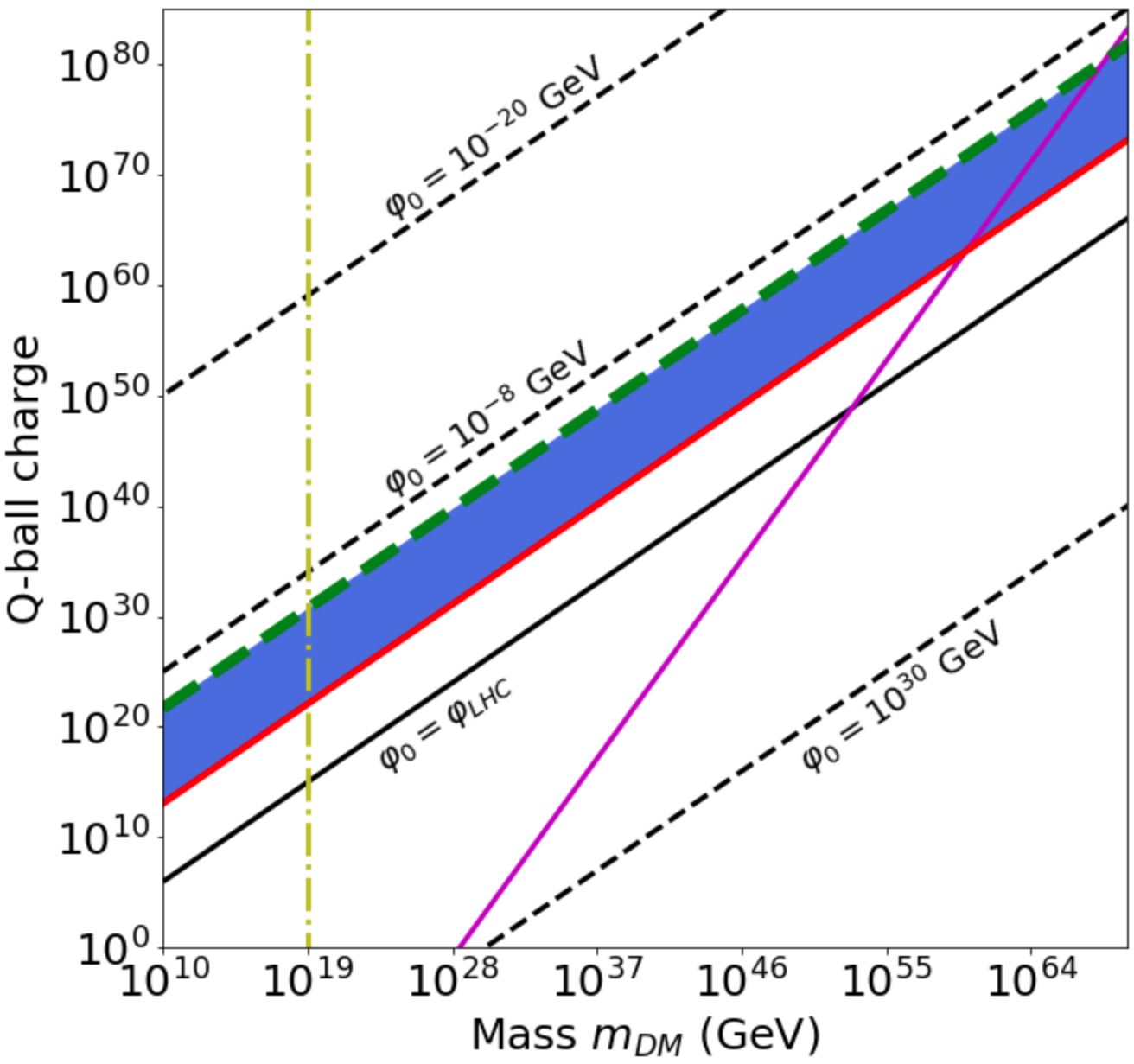}
    \caption{This is the same plot as Fig.~\ref{fig:Type3}, with the added magenta line representing the black hole limit as a minimum charge as a function of mass.}
    \label{fig:plot 3}
\end{figure}

We show in figures~\ref{fig:plot 1}-\ref{fig:plot 3} the extended parameter space of thick and thin-walled Q-balls. This is to include the regions where the black hole limits become relevant. Here we have added a magenta line representing the black hole constraints of each model. The black hole limit is a maximum mass constraint for a given charge indicating an allowed parameter space above the line. In all three cases the parameter space where the Q-ball is visible and point-like lies within the allowed regions. Therefore, point-like Q-balls that are visible to arrays of opto-mechanical sensors such as Windchime are not in danger of gravitational collapse.

\end{document}